\documentclass[reprint,superscriptaddress,amssymb,amsmath,aps,prx,longbibliography, nofootinbib]{revtex4-1} 

\usepackage{graphicx}
\usepackage{amssymb}
\usepackage{epstopdf}
\usepackage{upgreek}
\usepackage{dcolumn}
\usepackage{bm}
\usepackage{gensymb}
\usepackage{braket}
\usepackage{amsmath}
\usepackage{mathtools}
\usepackage{float}
\usepackage{tabularx}

\usepackage{tikz}

\expandafter\ifx\csname package@font\endcsname\relax\else
 \expandafter\expandafter
 \expandafter\usepackage
 \expandafter\expandafter
 \expandafter{\csname package@font\endcsname}%
\fi

\newcommand{\be}{\begin{eqnarray}}
\newcommand{\ee}{\end{eqnarray}}
\newcommand{\bfig}{\begin{figure}}
\newcommand{\efig}{\end{figure}}

\newcommand{\gsim}{\raisebox{-0.13cm}{~\shortstack{$>$ \\[-0.07cm]
      $\sim$}}~}
\newcommand{\lsim}{\raisebox{-0.13cm}{~\shortstack{$<$ \\[-0.07cm]
      $\sim$}}~}

\DeclareFontFamily{U}{mathb}{}
\DeclareFontShape{U}{mathb}{m}{n}{
  <-5.5> mathb5
  <5.5-6.5> mathb6
  <6.5-7.5> mathb7
  <7.5-8.5> mathb8
  <8.5-9.5> mathb9
  <9.5-11.5> mathb10
  <11.5-> mathbb12
}{}

\usepackage[colorlinks=true,allcolors=blue]{hyperref}%

\begin{document}

\title{An improved analysis framework for axion dark matter searches}
\author{D. A. Palken}
\email{daniel.palken@colorado.edu}
\affiliation{JILA, National Institute of Standards and Technology and the University of Colorado, Boulder, Colorado 80309, USA}
\affiliation{Department of Physics, University of Colorado, Boulder, Colorado 80309, USA}

\author{B. M. Brubaker}
\affiliation{JILA, National Institute of Standards and Technology and the University of Colorado, Boulder, Colorado 80309, USA}
\affiliation{Department of Physics, University of Colorado, Boulder, Colorado 80309, USA}
\affiliation{Department of Physics, Yale University, New Haven, Connecticut 06511, USA}

\author{M. Malnou}
\affiliation{JILA, National Institute of Standards and Technology and the University of Colorado, Boulder, Colorado 80309, USA}
\affiliation{Department of Physics, University of Colorado, Boulder, Colorado 80309, USA}

\author{S. Al Kenany}
\affiliation{Department of Nuclear Engineering, University of California Berkeley, California 94720, USA}

\author{K. M. Backes}
\affiliation{Department of Physics, Yale University, New Haven, Connecticut 06511, USA}

\author{S. B. Cahn}
\affiliation{Department of Physics, Yale University, New Haven, Connecticut 06511, USA}

\author{Y. V. Gurevich}
\affiliation{Department of Physics, Yale University, New Haven, Connecticut 06511, USA}

\author{S. K. Lamoreaux}
\affiliation{Department of Physics, Yale University, New Haven, Connecticut 06511, USA}

\author{S. M. Lewis}
\affiliation{Department of Nuclear Engineering, University of California Berkeley, California 94720, USA}

\author{R. H. Maruyama}
\affiliation{Department of Physics, Yale University, New Haven, Connecticut 06511, USA}

\author{N. M. Rapidis}
\affiliation{Department of Nuclear Engineering, University of California Berkeley, California 94720, USA}

\author{J. R. Root}
\affiliation{Department of Nuclear Engineering, University of California Berkeley, California 94720, USA}

\author{M. Simanovskaia}
\affiliation{Department of Nuclear Engineering, University of California Berkeley, California 94720, USA}

\author{T. M. Shokair}
\affiliation{Department of Nuclear Engineering, University of California Berkeley, California 94720, USA}

\author{Sukhman Singh}
\affiliation{Department of Physics, Yale University, New Haven, Connecticut 06511, USA}

\author{D. H. Speller}
\affiliation{Department of Physics, Yale University, New Haven, Connecticut 06511, USA}

\author{I. Urdinaran}
\affiliation{Department of Nuclear Engineering, University of California Berkeley, California 94720, USA}

\author{K. van Bibber}
\affiliation{Department of Nuclear Engineering, University of California Berkeley, California 94720, USA}

\author{L. Zhong}
\affiliation{Department of Physics, Yale University, New Haven, Connecticut 06511, USA}

\author{K. W. Lehnert}
\affiliation{JILA, National Institute of Standards and Technology and the University of Colorado, Boulder, Colorado 80309, USA}
\affiliation{Department of Physics, University of Colorado, Boulder, Colorado 80309, USA}
\date{\today}

\begin{abstract}
In experiments searching for axionic dark matter, the use of the standard threshold-based data analysis discards valuable information. We present a Bayesian analysis framework that builds on an existing processing protocol \cite{brubaker2017haystac} to extract more information from the data of coherent axion detectors such as operating haloscopes. The analysis avoids logical subtleties that accompany the standard analysis framework and enables greater experimental flexibility on future data runs. Performing this analysis on the existing data from the HAYSTAC experiment, we find improved constraints on the axion-photon coupling $g_\gamma$ while also identifying the most promising regions of parameter space within the $23.15$--$24.0$ $\mu$eV mass range. A comparison with the standard threshold analysis suggests a $36\%$ improvement in scan rate from our analysis, demonstrating the utility of this framework for future axion haloscope analyses. 

\end{abstract}

\maketitle

\section{INTRODUCTION} \label{sec:intro}

Dark matter axion detection experiments generically entail searching for a weak signal at an unknown frequency $\nu_a = m_ac^2/h$, where the axion mass $m_a$ is a free parameter whose possible values span many orders of magnitude \cite{graham2015experimental}. A near-optimal search strategy \cite{chaudhuri2018fundamental} requires a resonant detector whose frequency is tuned slowly over a range much larger than its bandwidth, with the spectral scan rate being the figure of merit \cite{malnou2019squeezed}. The technology enabling axion detection is most mature at rf and microwave frequencies \cite{malnou2019squeezed, much1998radio, castellanos2007widely, malnou2018optimal, kindel2016generation, johnson2010quantum, zheng2016accelerating, stern2015cavity}, where haloscope \cite{sikivie1985detection} experiments such as ADMX \cite{hagmann1998results, asztalos2002experimental, asztalos2004improved, asztalos2010squid, asztalos2001large, sloan2016limits, du2018search, braine2020extended, boutan2018piezoelectrically, crisosto2019admx} and HAYSTAC \cite{brubaker2017firstA, alkenany2017design, brubaker2017haystac, brubaker2017first, zhong2018results}, have achieved sensitivity to the QCD axion model band; ADMX is now sensitive to the Dine-Fischler-Srednicki-Zhitnitsky (DFSZ) benchmark coupling \cite{dine1981simple, zhitnitsky1980on, du2018search, braine2020extended}. If DFSZ sensitivity is necessary to detect the axion, the scaling of a single-cavity haloscope's scan rate $R$ with $\nu_a$ as approximately $R \propto \nu_a^{-14/3}$ \cite{brubaker2017first} becomes particularly foreboding. It implies that HAYSTAC and ADMX would each require roughly 20 thousand years to scan the $1$--$10$ GHz frequency decade, generously assuming noise at the quantum limit and $100\%$ live-time \cite{malnou2019squeezed}. It is therefore important to introduce improved detection and analysis protocols applicable across the full landscape of axion detector platforms \cite{kahn2016broadband,ouellet2018first, foster2018revealing, silvafeaver2017design, chaudhuri2018fundamental, mcallister2017organ, caldwell2017madmax, rybka2015search, lee2020axion, alesini2019galactic}. 

An optimization of the statistical analyses that haloscopes have historically used is readily available. The standard threshold-based confidence tests used in haloscope analyses transform continuous measurements of power into binary outcomes, discarding knowledge of what power was measured, and, with it, sensitivity to the axion. Because haloscopes are in practice statistics-limited experiments, for which additional data continues to enhance sensitivity, a more informative analysis will translate into tangible time-savings during operation. 

In this article, we introduce an analysis framework which simplifies the operational constraints placed on experimentalists while making use of more of the information content of coherent axion detection data. This framework builds on the existing HAYSTAC processing procedure \cite{brubaker2017haystac} and may be readily adapted to other experiments. We begin in Sec.\,\ref{sec:FT} by reviewing the standard threshold analysis, devoting particular attention to its discarding of valuable information and the difficulty of adhering to its rigid logical requirements. In Sec.\,\ref{sec:BPM}, we present our new framework, which we refer to as Bayesian power measured (BPM). We test the BPM framework in Sec.\,\ref{sec:HAYSTAC} by reanalyzing the HAYSTAC phase 1 data, and find that the BPM analysis puts tighter constraints on the axion within the HAYSTAC window. In addition, it spotlights within the HAYSTAC dataset the group of frequencies ``least unlikely" to contain the axion. The most prominent among these, though still unlikely to contain an axion, nonetheless stands out relative to any other frequency within the scan range. 

\section{FREQUENTIST THRESHOLD FRAMEWORK} \label{sec:FT}
The majority of existing axion haloscope exclusion has been obtained using a thresholding framework against which the alternative framework presented here is directly compared. The purpose of this section is to contextualize our analysis framework by explaining what has historically been meant by exclusion.\footnote{The story is complicated by the fact that not all exclusion historically uses the framework described in this section. For an example of an alternative framework, see Ref.\,\cite{foster2018revealing}. A comparison to other, less common frameworks is beyond the scope of this work.}

Axion haloscopes place limits on the axion-photon coupling strength $g_\gamma$\footnote{Physically, $g_\gamma$ can take positive or negative values, but since only $g_\gamma^2$ contributes to the axion signal power, we use $g_\gamma$ to stand in for $|g_\gamma|$ throughout this article.} by repeatedly measuring the power decaying out of a tunable resonant cavity and statistically evaluating its consistency with an axion-induced excess. A typical search involves averaging power spectra obtained from the fluctuations of the quadrature observables for an integration time of order minutes, before tuning the apparatus to a nearby frequency and repeating. A processing protocol \cite{brubaker2017haystac} appropriately filters and combines these spectra.

\begin{figure*}[t] 
	\centering
	\includegraphics[scale=1.00]{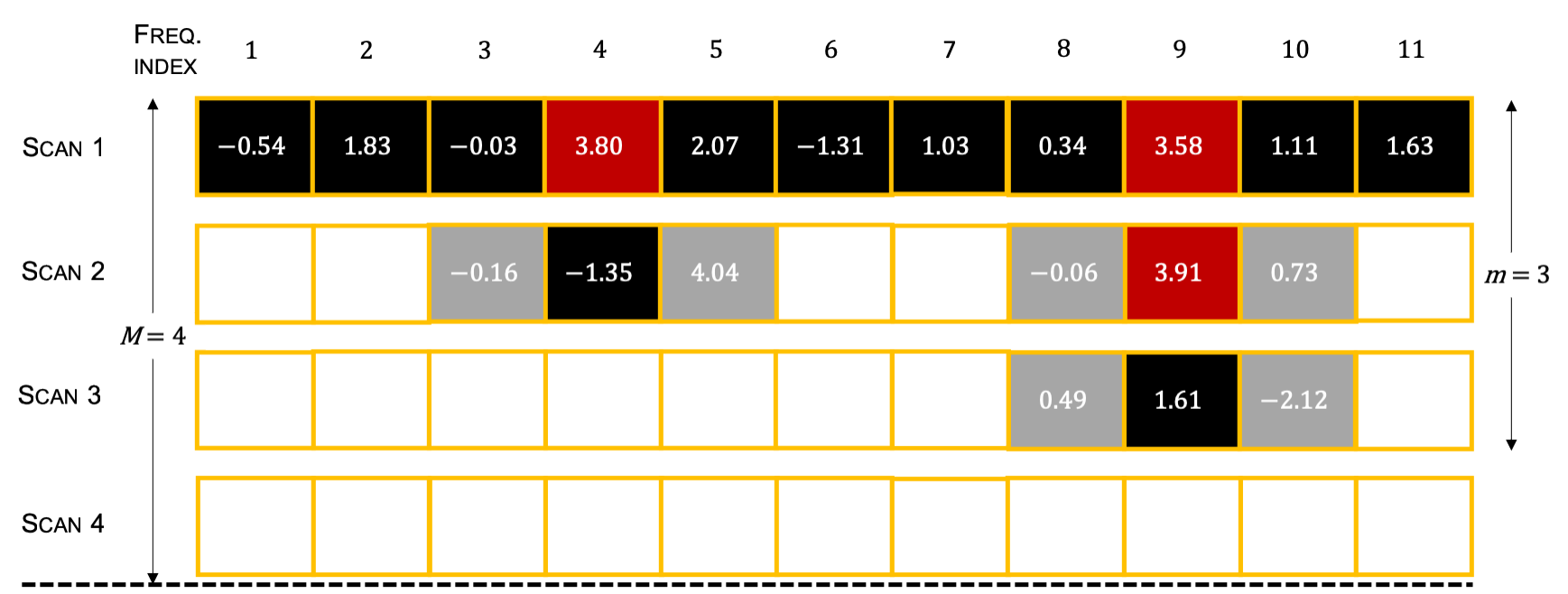} 
	\caption{Illustrative, fictitious grand spectrum data for a standard haloscope subject to a frequentist threshold (FT) analysis allowing $M=4$ scans. The initial scan data from a typical haloscope \cite{du2018search, zhong2018results} consists of a normalized power excess in each of $N \sim 10^5$--$10^6$ frequency bins, represented above by the boxes in the first row. At frequencies where the excess is above a predetermined threshold (red boxes), additional scans are conducted until a measurement below threshold is recorded. Only if all $M$ scans exceed threshold is failure to reject the axion hypothesis reported. If any scan comes in below threshold (black boxes), the axion hypothesis is rejected for the bin in accordance with Eq.\,\eqref{tot F neg} and further scanning need not be conducted (empty boxes). For the FT framework of Sec.\,\ref{sec:FT}, both the threshold and the number of scans $M$ must be predetermined, and the exclusion therefore may take into account scans that were not performed. Here, Scan 4 was not performed for any bins, but still impacts the final reported exclusion. Because the haloscope bandwidth is typically larger than the expected axion linewidth, bins adjacent to those exceeding threshold are automatically rescanned (gray boxes). However, the FT framework discards these data regardless of the power measured. Conversely, the BPM method discussed in Sec.\,\ref{sec:BPM} is able to use the information.} 
	\label{fig:Data Rep}
\end{figure*}

\begin{figure}[!h] 
	\centering
	\includegraphics[scale=0.5]{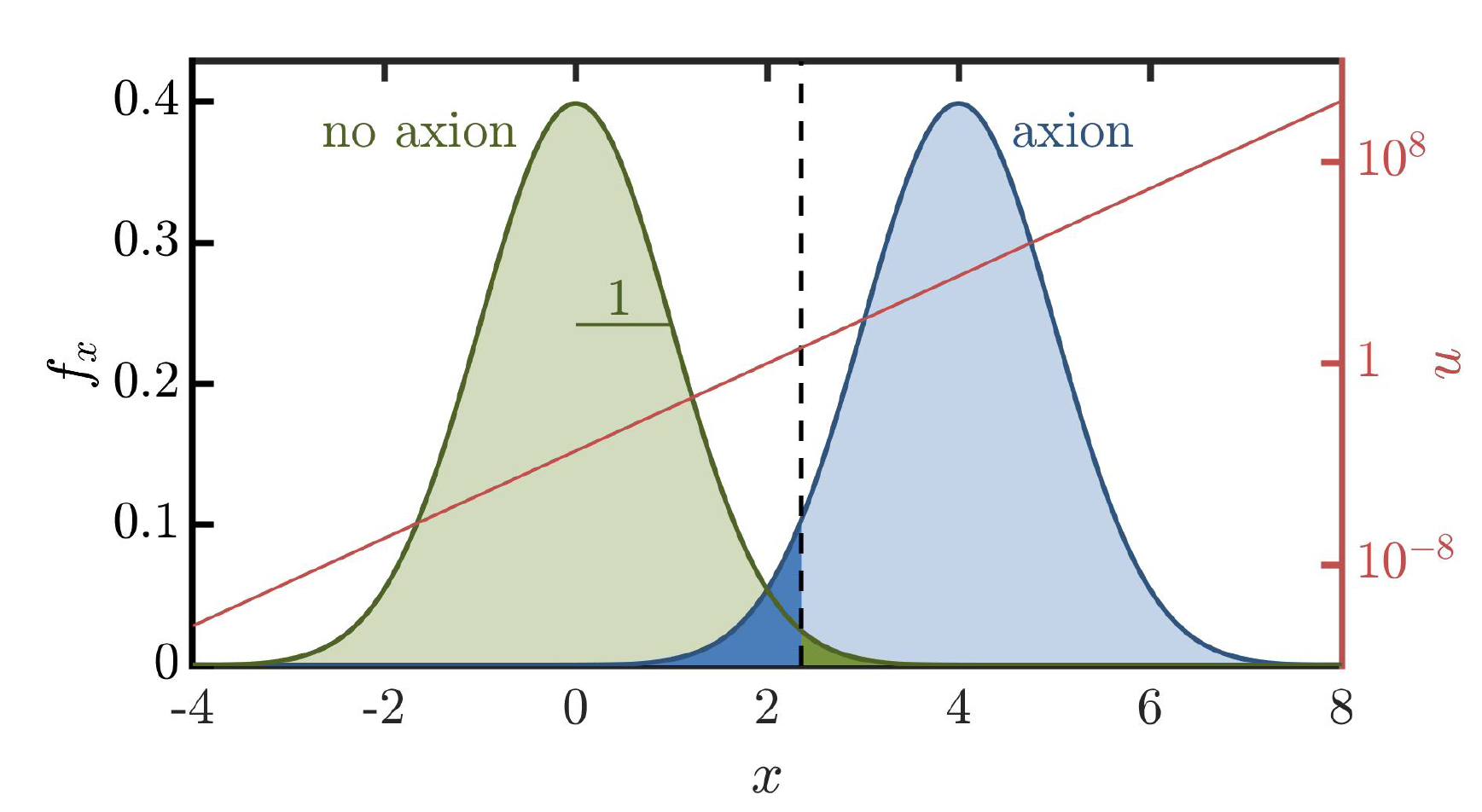}
	\caption{Gaussian probability density functions $f_x(x)$ for a single-bin normalized power excess $x$ owing to noise alone (green) and noise plus-axion signal for a particular value of the axion-photon coupling $g_\gamma$ and haloscope sensitivity (blue). In the low-prior limit, the single-scan prior update $u(x)$ (red) is simply the ratio of the two distributions. The exponential dependence of $u$ upon $x$ captures the information content of the measurement. In threshold-based inference frameworks, conversely, an excess power threshold (black dashed line) is predetermined, and the only information analyzed is whether the measurement is above threshold. The light green (blue) region denotes to a true-negative (-positive) indication, while the dark green (blue) denotes a false-positive (-negative).}
	\label{fig:2Gaus}
\end{figure}

This article starts where that data processing ends: with a grand spectrum of normalized power excesses predominantly from thermal fluctuations of the electromagnetic field. These normalized excesses $x^{(1)}_i$ are obtained by subtracting off the mean of the grand spectrum power distribution and dividing by its standard deviation. They are measured at frequencies $\nu_i$, with $i = 1, \cdots, N$ indexing the $N \gg 1$ frequency bins. The superscript $(1)$ denotes data taken on an initial scan. Bins displaying large excesses on an initial scan are rescanned, potentially multiple times, yielding rescan spectra $x^{(j)}_i$ indexed as $j = 2, \cdots, M$, where many frequency bins $i$ will not be measured in rescans. Figure \ref{fig:Data Rep} shows how a realistic grand spectrum's frequency bins might be populated with measured power excess data. For bins not containing an axion, acquisitions of order a millionfold times longer than the inverse bin bandwidth yield, via the central limit theorem, normalized power excesses drawn from the standard Gaussian probability density function (PDF), obtained by setting $\mu = 0,\ \sigma = 1$ in
\begin{equation} \label{pdf}
f_x(x; \mu, \sigma) = \frac{1}{\sqrt{2\pi}\sigma}\exp\left[-\frac{(x - \mu)^2}{2\sigma^2} \right].
\end{equation}
This distribution is shaded green in Fig.\,\ref{fig:2Gaus}. For a discussion of the consequences of practical departures from the Gaussian idealization, see Appendix \ref{app:non-Gauss}. 

A bin $\nu_i$ that falls at the axion frequency $\nu_a$ will have its power excess drawn from what we term the axion distribution. There is in fact a family of such distributions parametrized by the axion-photon coupling $g_\gamma$, but it will suffice to consider just one. The axion distribution's standard deviation in the limit of weak axion signal power is approximately that of the no-axion distribution (i.e. $1$) but its mean is offset to 
\begin{equation} \label{mu axion}
    \mu_{a,i}^{(j)} = g_{\gamma,i}^2 \eta_i^{(j)}. 
\end{equation}
The sensitivity parameter $\eta_i^{(j)}$ is the independently calibrated signal-to-noise ratio for an axion with coupling strength $g_\gamma = 1$ in the $i^\text{th}$ bin of the $j^\text{th}$ spectrum.\footnote{The parameters $\eta_i^{(j)}$ account for the scaling of the haloscope sensitivity with integration time, experimental parameters, and the dark matter density $\rho_a$. By normalizing the signal-to-noise ratio to $g_\gamma=1$ and treating the multiplicative coupling factor $g_\gamma^2$ as a parameter of the axion distribution, we are following the convention in the axion search literature of assuming a fixed value of $\rho_a$ and setting limits on $g_\gamma$. The $i$ index on the coupling factor indicates that we need not consider the same axion distribution in each bin.} The PDF of the axion distribution for the $i^\text{th}$ bin (blue in Fig.\,\ref{fig:2Gaus}) is thus Eq.\,\eqref{pdf} with $\mu = \mu_{a,i}^{(j)},\ \sigma = 1$. 

A power excess threshold $x^{(j)}_T$ is then set for each scan $j$ and this threshold is used to determine firstly whether or not to proceed with rescans, and secondly whether or not to exclude the axion hypothesis at each frequency. The logic of this exclusion is prescribed as follows. First, the experimentalist establishes a null hypothesis. In this case, the null hypothesis for each bin is that there \textit{is} an axion at the bin's frequency.\footnote{The axion hypothesis is chosen as the null so that it might be rejected, or excluded, later. Because of the unconventional choice to make the interesting (i.e. axion) hypothesis the null, the language of false-negatives and -positives will adopt the unconventional usage where a positive (negative) refers to the null being true (false).} Second, the experimentalist establishes and follows a procedure for acquiring data. This procedure may include conditional steps: e.g. ``acquire more data in a given bin via a rescan if and only if the initial scan power exceeds the threshold set for that bin." All possible paths of action eventually terminate either in a rejection of the null or a failure to reject the null, forming a decision tree. Exclusion is set for all bins $i$ where the null is rejected at a confidence level of 
\begin{equation}\label{excl}
    E = 1 - F_n,
\end{equation}
where $F_n$ is the bin-$i$ false-negative rate of the entire procedure: the total probability that the null would have been rejected, were it true. This approach is equivalent to defining a test statistic
\begin{equation}\label{test stat}
    t_s = \begin{cases}
    1 & x^{(j)} \geq x_T^{(j)}\, \forall\, j \\
    F_n(g_\gamma) & \text{otherwise} \\
    \end{cases}
\end{equation}
from the total false negative rate treated as an explicit function of $g_\gamma$. Couplings where $t_s$ goes below some predetermined target rate $F_{n,\text{target}}$ are rejected at $1-F_{n,\text{target}}$ confidence level via Eq.\,\eqref{excl}.\footnote{The piecewise definition of $t_s$, Eq.\,\eqref{test stat}, yields exclusion only when at least one scan's power excess $x^{(j)}$ comes in below its threshold $x_T^{(j)}$. When all scans exceed threshold, $t_s = 1$ indicates the presence of (possibly axionic) excess power.} We refer to this class of methods as frequentist threshold (FT) frameworks. FT frameworks answer the question, ``assuming the axion exists, what is the probability of failing to observe it?"

A FT search procedure \cite{zhong2018results} is prepared to perform as many as $M$ scans at each frequency $\nu_i$, reporting exclusion subject to its predetermined false-negative rate. At a given $\nu_i$, no additional scans are performed once any scan's measured power excess $x^{(j)}_i$ fails to exceed its predetermined threshold $x^{(j)}_T,$ the vertical, dashed line in Fig.\,\ref{fig:2Gaus}. For the general case where each scan may use a different false-negative rate (dark blue region of Fig.\,\ref{fig:2Gaus}), given from integrating Eq.\,\eqref{pdf} by
\begin{equation} \label{indiv F neg}
f_n^{(j)} = \int_{-\infty}^{x_T^{(j)}} f_x\left(x; \mu_a^{(j)},1\right) dx,
\end{equation}
the total false-negative rate is 
\begin{equation} \label{tot F neg}
    F_n = \sum_{j=1}^{M}f_n^{(j)}\prod_{k=1}^{j-1} \left(1-f_n^{(k)}\right).
\end{equation}
This total false-negative rate is what sets the confidence of the reported exclusion $E$ as in Eq.\,\eqref{excl}. For the special case where all $M$ scans have equal false-negative rates, $f_n^{(j)} = f_{n0},$ Eq.\,\eqref{tot F neg} reduces to
\begin{equation} \label{simple tot F neg}
    F_{n0} = 1 - (1 - f_{n0})^M.
\end{equation}
In practice, the false-negative rates $f_n^{(j)}$ and power thresholds $x_T^{(j)}$ are set to convenient values, removing the frequency-dependence of $\mu_a^{(j)} = g_{\gamma,i}^2\eta_i^{(j)}$ by forcing $g_\gamma$ to compensate for the frequency-dependence of the sensitivity. The initial scan sensitivities $\eta_i^{(1)}$ therefore determine the minimum coupling $g_{\gamma,i}$ that is excluded (or not) at each frequency. Subsequent ($j>1$) scans, conversely, integrate to a sensitivity $\eta_{i}^{(j)}$ determined by the coupling $g_{\gamma,i}$ reached on the initial scan. The result, in the case where the axion hypothesis is rejected for all bins where it can be tested, is an exclusion plot \cite{zhong2018results} whose spectral structure reflects that of $\eta_i^{(1)}$. 

The procedure for collecting data and reporting exclusion in these FT frameworks must be rigidly defined and laid out in advance. If, instead, the experimentalist is allowed to alter the decision tree that leads to negative or positive results while navigating that decision tree, the result becomes susceptible to bias. To this point, the total false-negative rate $F_n$, Eq.\,\eqref{tot F neg}, directly responsible for setting the exclusion $E$, grows with the total number $M$ of scans in the predetermined experimental procedure. Suppose, then, that the experimentalist was willing to perform as many as $M$ scans at each frequency bin $i$ before failing to reject the axion hypothesis, but that in practice some number $m < M$ was all that was required to get at least one negative result in every bin. The correct exclusion to report, perhaps unintuitively, is that which takes into account all $M$ allowed scans, including even the $M - m$ unperformed scans. The number of such unperformed scans and their individual false negative rates $f_n^{(j)}$, Eq.\,\eqref{indiv F neg}, must be known to properly report exclusion. If not written down in advance, these numbers are difficult to estimate without inserting bias. In practice it is extremely difficult to rigorously adhere to FT logic. For example, one would expect that some exclusion would be reported for a total number of scans $M$ surpassing $m$, the exact number performed, yet this never appears in the literature \cite{depanfilis1987limits, wuensch1989results, hagmann1990results, hagmann1990search, hagmann1998results, asztalos2002experimental, asztalos2004improved, asztalos2010squid, sloan2016limits, brubaker2017firstA}. Furthermore, exclusion has been reported for a number of scans less than the number performed \cite{brubaker2017haystac}. 

Though the preceding logic and inference is frequentist in nature, an axion haloscope analysis can alternatively be carried out using the language of Bayesian statistics. In Appendix \ref{app:BT}, we describe a threshold-based Bayesian analysis (denoted BT2) which is equivalent to the FT framework, given simple assumptions applicable to all dark matter axion experiments. That is, the two frameworks share operational procedures, and they ultimately output identical conditional probabilities. This correspondence implies that we can quantitatively interpret FT axion exclusion as decreased probability of the axion hypothesis. 

Adopting a Bayesian perspective, we can now ask whether a more informative analysis is possible. We consider frameworks which operate by applying Bayes' theorem,
\begin{equation} \label{Bayes}
    P(Y|Z) = \frac{P(Z|Y)P(Y)}{P(Z)},
\end{equation}
to an axion search data set \cite{wainstein1962detection, harman1963principles}. Notationally, $P(B)$ denotes the probability of event $B$ being true, while $P(B|C)$ denotes the same, conditional upon event $C$ being true. In Eq.\,\eqref{Bayes}, $P(Y)$ is the Bayesian prior probability of event $Y$ being true, and it is updated by the occurrence of event $Z$ to the posterior probability $P(Y|Z)$. Appendix \ref{app:BT} describes two distinct Bayesian threshold (BT) frameworks in which $Z$ is taken to be a binary outcome, or set thereof. We show that a more informative analysis is indeed possible, but that thresholding always imposes some operational restrictions and needlessly discards valuable information. Nonetheless, the BT logic exemplifies how Bayes' theorem may be applied to axion detection. In doing so, it builds a bridge between the threshold methods of this section and the BPM framework of the next.

\section{BAYESIAN POWER-MEASURED FRAMEWORK} \label{sec:BPM}
Bayes' theorem, Eq.\,\eqref{Bayes}, may be applied to haloscope data so as to preserve the information content of the measurement. The information at each frequency may be aggregated into a statement about the change in probability of an axion of arbitrary coupling strength $g_{\gamma,i}$ existing anywhere within the haloscope's scan range. In the common case where the data indicates the absence of an axion, the Bayesian language of updated probability maps onto the language of frequentist exclusion (Appendices \ref{app:BT} and \ref{app:agg excl}, and Ref.\,\cite{lista2017statistical}). In this section, we discuss the BPM framework, and how its posteriors can be aggregated, before applying it to the HAYSTAC phase 1 data in Sec.\,\ref{sec:HAYSTAC}. We address potential concerns about the inevitable non-Gaussianity of real haloscope data in Appendix \ref{app:non-Gauss}, about the seemingly subjective choice of priors in Appendix \ref{app:priors}, and about the information content of the grand spectrum data in Appendix \ref{app:combine}.

The BPM framework is readily motivated by the fact that both the FT (Sec.\,\ref{sec:FT}) and BT (Appendix \ref{app:BT}) frameworks discard valuable information. Each power excess, measured on a continuum, is reduced to a simple ``click" or ``no-click" response of an effective binary detector. A measurement that comes in just below threshold on a higher-order rescan after already exceeding threshold on previous scans is orders of magnitude more likely to indicate the presence of an axion than a typical, $0\sigma$ power excess. Yet the threshold frameworks treat these events identically. 

The BPM framework uses Bayes' theorem to account for the precise effect of any initial measured power $x_i^{(1)}$ on the prior probability $P_{a,i}^{(0)}$ that an axion resides in bin $i$. After measurement, the prior is updated to the posterior, $P_{a,i}^{(1)} = u_i^{(1)} P_{a,i}^{(0)}$, where $u_i^{(1)}$ denotes the first scan's prior update. If subsequent scans $j>1$ are performed, $P_{a,i}^{(j-1)}$ will be further updated to $P_{a,i}^{(j)} = u_i^{(j)}P_{a,i}^{(j-1)}$, and so forth. In the appropriate limit of infinitesimal priors (Appendix \ref{app:priors}), the single-scan prior update for bin $i$ and scan $j$ follows from Eq.\,\eqref{Bayes}, independent of one's choice of prior:
\begin{equation} \label{BPM update}
    u_i^{(j)} \approx \frac{P\left(x_i^{(j)}|\mathcal{A}_i\right)}{P\left(x_i^{(j)}|\mathcal{N}_i\right)} = \exp\left[-\frac{\left(\mu_{a,i}^{(j)}\right)^2}{2} +\mu_{a,i}^{(j)} x_i^{(j)}\right].
\end{equation}
The second equality uses the Gaussian probability densities of Eq.\,\eqref{pdf}, and $\mathcal{A}_i$ or $\mathcal{N}_i$ denotes the event that an axion does or does not reside in bin $i$, respectively. The approximation in Eq.\,\eqref{BPM update} identifies the prior update with the Bayes factor, which compares the likelihood of two hypotheses; it applies when the axion and other sources of excess power are sufficiently unlikely. The assumption underlying the validity of this approximation is  made quantitative in Appendix \ref{app:priors}, and in Appendix \ref{app:non-Gauss} we show that this assumption is conservative with respect to axion exclusion. The BPM analysis can therefore alternatively be viewed as a Bayes factor analysis, but the interpretation of the Bayes factor as a conservative approximation to the prior update makes for a more intuitive and immediately useful end result. Equation \eqref{BPM update} also indicates that the update to the probability of an axion's existence, plotted in red in Fig.\,\ref{fig:2Gaus}, scales exponentially with the measured power excess $x_i^{(j)}$. Rather than assume a particular axion distribution, we now consider the prior update $u_i^{(j)}$ as a function of $\mu_{a,i}^{(j)}$. Formally, $u_i^{(j)}$ is maximized at $\mu_{a,i}^{(j)}=x_i^{(j)}$. Thus at each frequency where a positive excess $x_i^{(j)} > 0$ is measured, we can single out a maximum likelihood (ML) axion distribution parametrized by $\mu_{a,i}^{(j)}=x_i^{(j)}$.\footnote{Conversely, measurement of a negative excess $x_i^{(j)}<0$ reduces the probability of any axion relative to the no-axion hypothesis.} The update to the prior probability in the ML axion scales sharply with $\mu_{a,i}^{(j)} \propto g_\gamma^2$ as $u_{i,\text{ML}}^{(j)} = \exp\left[(\mu_{a,i}^{(j)})^2/2 \right]$. Since the updates $u_i^{(j)}$ can be mapped back to their measured power excesses $x_i^{(j)}$ and sensitivities $\eta_i^{(j)}$, this use of Bayes' theorem preserves the information content of each individual measurement. 

Applying Eq.\,\eqref{BPM update} to all bins $x_i^{(j)}$ across all scans yields a spectrum of total updates $U_i$, defined as
\begin{equation} \label{BPM update tot}
    U_i = \frac{P_{a,i}^{(M_i)}}{P_{a,i}^{(0)}} = \prod_{j=1}^{M_i} u_i^{(j)},
\end{equation}
    where $M_i$ is the total number of scans performed on bin $i$. Unlike in the FT framework of Sec.\,\ref{sec:FT}, scans that are not performed do not enter into the equation, and the number of scans need not be specified in advance. Subject to its equivalence to the Bayes factor (Appendices \ref{app:non-Gauss} and \ref{app:priors}), each total update $U_i$ is the change in probability that an axion of a specified coupling $g_\gamma$ resides in the bin. For example, $U_i = 0.1$ implies that it is $10\%$ as probable that an axion resides in the bin after all $M_i$ scans were performed than it was prior to the experiment. A total prior update of $U_i = 0.1$ carries exactly the same meaning as $90\%$ exclusion ($E = 0.9$) in the standard analysis framework, in the limit of vanishing false-positive rates (Appendix \ref{app:BT}). 

\begin{figure*}[t] 
	\centering
	\includegraphics[scale=1.00]{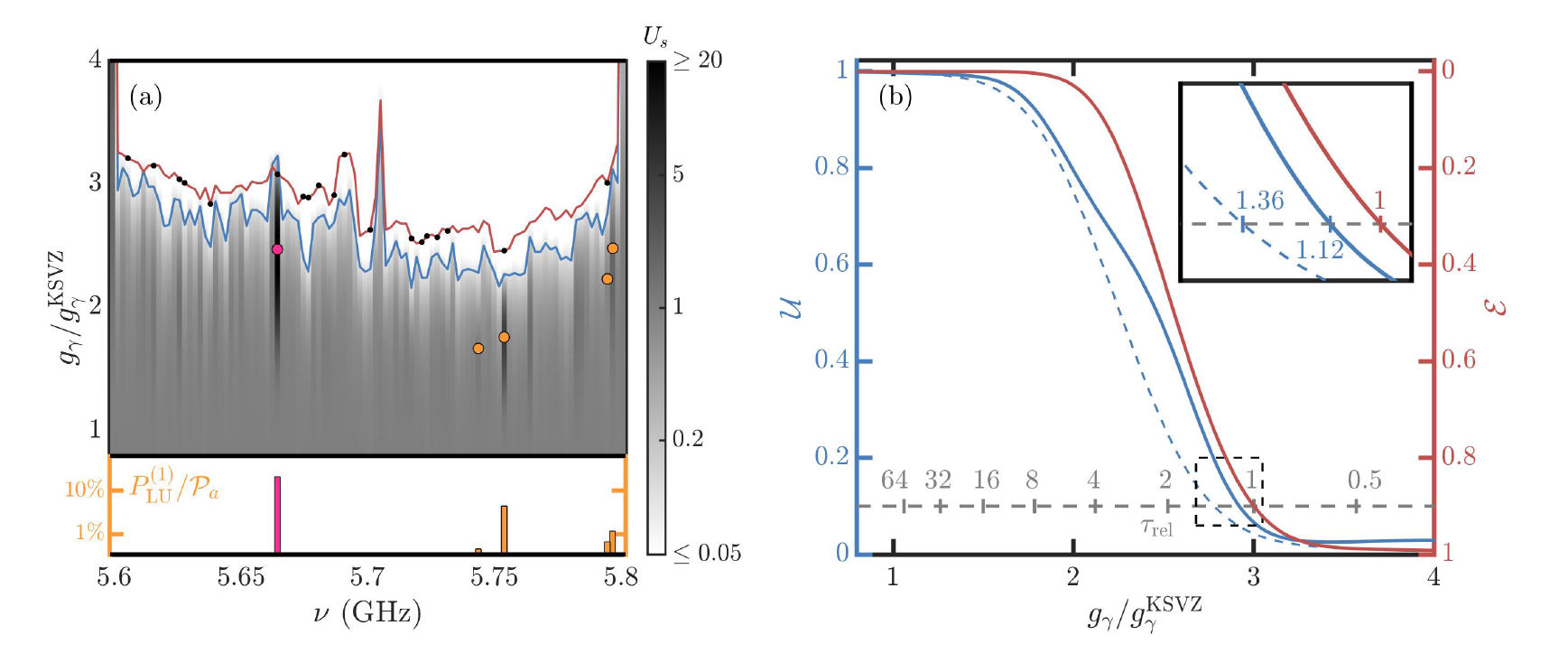} 
	\caption{(a) Reanalysis of HAYSTAC phase 1 data as a test of the Bayesian power measured (BPM) framework. The standard $90\%$ exclusion line (red) achieved with the FT framework \cite{zhong2018results} is equivalent to the $10\%$ prior update contour of the second Bayesian threshold (BT2) framework discussed in Appendix \ref{app:BT}. The $10\%$ prior update contour (blue) achieved with the BPM analysis constrains more aggressive couplings at nearly all frequencies. The logarithmic color scale indicates the subaggregated prior updates $U_s$ throughout the full two-dimensional parameter space. Darker (lighter) shading corresponds to increased (decreased) probability of an axion of given frequency $\nu$ and axion-photon coupling $g_\gamma$ existing. Frequencies where rescans were performed are marked with a black dot on the threshold exclusion line. The five least unlikely (LU) axion candidates --- those whose priors increased most --- are marked at their maximum likelihood couplings with orange circles. In the lower panel, corresponding orange bars indicate on a logarithmic axis the total posterior probability $P_\text{LU}^{(1)}$ within the BPM framework that an axion exists at each of the LU points normalized to the aggregate prior probability $\mathcal{P}_a$ that an axion resides anywhere in the window at that coupling. The LU candidate at $\nu_\text{LU}^\star = 5.66417255\ \text{GHz}$, marked in pink, is more than two times higher than the remainder of the top five combined. (b) Aggregate prior update $\mathcal{U}$ (equivalently, exclusion $\mathcal{E}$) taken as a function of coupling $g_\gamma$ over the entire HAYSTAC phase 1 frequency window for the FT and BT2 frameworks (solid red line), BPM framework (solid blue line), and BPM framework with several adjustments made for a clean comparison with the performance of the thresholding (dashed blue line; see main text). The dashed, gray line, zoomed in upon in the inset, indicates relative $\mathcal{U} = 10\%$ ($\mathcal{E} = 90\%$) scan time normalized to that achieved with the FT and BT2 frameworks \cite{zhong2018results}. Under typical experimental conditions, the exclusion achievable with the BPM framework would take an estimated $36\%$ longer to attain using thresholding.}
	\label{fig:HAYSTAC data}
\end{figure*}

The individual $P_{a,i}^{(j)}$, or $U_i$, regarded as functions of $g_\gamma$, can be spectrally combined into an aggregate prior update\footnote{In this text, we refer to four kinds of prior update. The \textit{single-scan} updates $u_i^{(j)}$ of Eq.\,\eqref{BPM update} are multiplicatively combined into the \textit{total} updates of Eq.\,\eqref{BPM update tot} at each frequency. The total updates are subsequently aggregated via Eq.\,\eqref{agg update} into the \textit{aggregate} update. The \textit{subaggregated} updates of Fig.\,\ref{fig:HAYSTAC data}a are likewise aggregated, but over $1\%$ of the HAYSTAC search window apiece. All updates are functions of axion-photon coupling $g_\gamma$, and are written explicitly as such where relevant. The quantities as calculated are Bayes factors, identifiable as updates subject to Appendices \ref{app:non-Gauss} and \ref{app:priors}.} function:
\begin{equation} \label{agg update}
    \mathcal{U}(g_\gamma) = \frac{\mathcal{P}_a^\prime(g_\gamma)}{\mathcal{P}_a(g_\gamma)} = \frac{\sum_{i=1}^N P_{a,i}^{(M_i)}(g_\gamma)}{\sum_{i=1}^NP_{a,i}^{(0)}(g_\gamma)}
    \approx
    \frac{1}{N}\sum_{i=1}^N U_i(g_\gamma),
\end{equation}
where $\mathcal{P}_a^\prime(g_\gamma)$ and $\mathcal{P}_a(g_\gamma)$ are the aggregate posterior and prior probabilities, respectively, of that axion existing. The aggregate update $\mathcal{U}(g_\gamma)$ does for the entire haloscope run what the total update $U_i(g_\gamma)$ does for each bin: it expresses the update to the probability that an axion of coupling $g_\gamma$ resides anywhere within the scanned frequency window. The final equality of Eq.\,\eqref{agg update} holds in the limit of approximately uniform priors, $P_{a,i}^{(0)} \approx \mathcal{P}_a / N$, applicable when the total frequency scan range is small compared to any of the scan frequencies, $\nu_N - \nu_1 \ll \nu_1$, as is the case for HAYSTAC \cite{zhong2018results} or ADMX \cite{du2018search}. For low-frequency experiments such as ABRACADABRA \cite{kahn2016broadband} or DM Radio \cite{silvafeaver2017design}, logarithmically uniform priors, Eq.\,\eqref{log priors}, should be used  \cite{chaudhuri2018fundamental}. 

Regardless of the measured excess powers $x_i^{(j)}$ and sensitivities ${\eta_i^{(j)}}$, there will exist a coupling $g_{\gamma,\text{low}}$ below which virtually nothing is learned about the presence or absence of an axion, $\mathcal{U}(g_\gamma \leq g_{\gamma,\text{low}}) \approx 1$. Likewise, there will exist another coupling $g_{\gamma,\text{high}}$ above which the probability of an axion existing vanishes $\mathcal{U}(g_\gamma \geq g_{\gamma,\text{high}}) \approx 0$. Between these two extremes, $\mathcal{U}(g_\gamma)$ will depend strongly on the measured power spectra. In the context of a realistic search, for which the experimental overhead is high, $\mathcal{U} > 1$ suggests that at least some bins within the range should be rescanned until either a cause of excess power is found or $\mathcal{U}$ regresses to a lower value. 

The aggregate prior update has several features that account for the look-elsewhere effect --- the linear growth in expected number of excesses of a given, otherwise significant size with number of independent hypotheses tested (for dependent hypotheses, see Appendix \ref{app:corr}). Because the denominator of Eq.\,\eqref{agg update} also grows linearly with tests performed, these two linear factors cancel out. Secondly, the typical coupling excluded will move upwards (Appendix \ref{app:expected outcomes}) as more independent tests are conducted, reflective of the increasing difficulty in ruling out a special frequency when it has more imposters to hide among. Finally, the ability to incorporate nonuniform priors is a key feature for experiments sensitive over fractionally large spectral windows. Since $90\%$ of all independent axion tests in such an experiment will occur in the highest decade, a standard frequentist trials factor approach \cite{lista2017statistical} would set an artificially higher bar for discovery in lower decades than it would have if the highest decade were not scanned; logarithmic priors scale the height of that bar with the number of tests performed. 

\section{TEST USING HAYSTAC PHASE 1 DATASET} \label{sec:HAYSTAC}

In order to test the BPM framework, we reanalyzed the full dataset from phase 1 of the HAYSTAC experiment, which includes at several frequencies as many as $m = 3$ scans. Features specific to this dataset are discussed in Ref.\,\cite{zhong2018results}. Notably, several frequency bins in the original analysis were discarded for having excess rf power that could with certainty be identified as at least partially nonaxionic in nature. We set $U_i = 1$ for these bins. Otherwise, straightforward application of Eqs.\,\eqref{BPM update tot} and \eqref{agg update} yield the results in Figs.\,\ref{fig:HAYSTAC data}a and b, respectively.  

The heat-map of Fig.\,\ref{fig:HAYSTAC data}a shows the updates to the probability of the axion existing throughout the two-dimensional parameter space. At high couplings ($g_\gamma \geq g_{\gamma,\text{high}} \approx 3g_\gamma^\text{KSVZ}$),\footnote{The benchmark Kim-Shifman-Vainshtein-Zakharov (KSVZ) \cite{kim1979weak, shifman1980can} axion has coupling $g_\gamma^\text{KSVZ} = 0.97$, making it a far more optimistic benchmark than the DFSZ axion, with $g_\gamma^\text{DFSZ} = 0.36$.} the probability of an axion is greatly diminished from what it was prior to collecting data. At low couplings ($g_\gamma \leq g_{\gamma,\text{low}}\approx g_\gamma^\text{KSVZ}$), virtually no information can been gleaned from the data about the presence or absence of an axion. At intermediate couplings, $g_{\gamma,\text{low}} < g_\gamma < g_{\gamma,\text{high}}$, updates to the probability of an axion being present depend sharply on the specific measured powers $x_i^{(j)}$, and also on the somewhat smoother haloscope sensitivity profile $\eta_i^{(j)}$. 

The FT framework has its $90\%$ exclusion line plotted in red in Fig.\,\ref{fig:HAYSTAC data}a. This corresponds directly\footnote{The correspondence comes with the caveat that the data here has been subaggregated into bins of width $\approx 2$ MHz, narrower than would be of practical concern to any experimentalist considering a haloscope run at nearby frequencies.} to the exclusion plotted in Ref.\,\cite{zhong2018results}, and equivalently to the $10\%$ prior update contour of the BT2 framework of Appendix \ref{app:BT}. The solid, blue line is the BPM framework's equivalent $10\%$ prior update contour. At almost all frequencies, it constrains more pessimistic couplings than do the threshold frameworks. 

At the few frequencies where the BPM framework indicates weaker exclusion than the threshold frameworks, it does so with good reason. Most of these ``least unlikely" (LU) axion candidate frequencies had excess power above threshold on an initial scan and not far below threshold on a rescan. This can be seen from the spectral coincidence of several of these LU candidates, marked with orange dots at their ML couplings, with the black dots indicating the performance of at least one rescan. The LU candidates that were not rescanned corresponded to initial scan powers coming in just below threshold. 

In the bottom panel of Fig.\,\ref{fig:HAYSTAC data}a, the prior updates for the five LU candidates are plotted normalized to the aggregate prior probability that an axion resides in the HAYSTAC phase 1 window. As discussed in Ref.\,\cite{chaudhuri2018fundamental}, it is not obvious how to distribute one's prior probability of an axion's existence with respect to coupling $g_\gamma$. However all five LU candidates have ML coupling strengths within a factor-of-1.5 range, wherein priors should not differ greatly. Therefore, the \textit{relative} significance of the highest of these LU candidate prior updates (frequency $\nu_\text{LU}^\star = 5.66417255\ \text{GHz}$) is noteworthy: it is more than two times higher than the remaining LU candidates combined. 

It should be emphasized that despite its exceptionally high prior update relative to the rest of the dataset, it is overwhelmingly unlikely that this LU candidate is an axion. A rough attempt at estimating priors (Appendix \ref{app:priors}) puts an optimistic probability of an axion in this bin still below part-in-1,000, with more realistic estimates well below that. Secondly, as discussed in Appendix \ref{app:non-Gauss}, the BPM framework is inherently conservative with respect to exclusion in a real dataset, and consequently liberal in its identification of candidates. In particular, nonaxionic rf power excesses caused other spikes in the HAYSTAC phase 1 power spectra. Many of these were manually removed from the dataset on the grounds of supplementary evidence (e.g. failure to persist when the auxiliary, weakly-coupled antenna was pulled out of the cavity) \cite{zhong2018results}. Since this LU candidate came in under threshold on a third and final scan, it was rejected subject to the FT framework. However, the more informative BPM framework suggests that significantly more than any of the other HAYSTAC phase 1 frequencies, it merits further interrogation. Generously precluding the possibility of nonaxionic excess power contamination, its probability of containing an axion at its maximum likelihood coupling is roughly the probability of $40$ MHz of nearby, unscanned parameter space, or $P_\text{LU}^{(1)} / \mathcal{P}_a = 20\%$ of the $200$ MHz scan window, containing an axion at that coupling. 

The frequency-dependent update (or exclusion) data of Fig.\,\ref{fig:HAYSTAC data}a has been aggregated in accordance with Eq.\,\eqref{agg update} and Eq.\,\eqref{agg excl} for the BPM and FT frameworks, respectively, in Fig.\,\ref{fig:HAYSTAC data}b. The solid, red line indicates aggregate updated probability (left axis) or exclusion of (right axis) an axion anywhere within the HAYSTAC phase 1 frequency window as a function of coupling. The solid, blue line is the updated probability for the BPM framework. Its upward bulge around $2.5g_\gamma^\text{KSVZ}$ is largely due to the LU candidate already discussed. Had the knowledge from the BPM framework been in hand before the HAYSTAC phase 1 decommissioning \cite{brubaker2017first}, little additional scan time would have been required to either identify or promptly rule out an axion signal. Therefore, an exclusion curve more representative of the BPM framework for comparison purposes is given by the dashed, blue curve, where the prior update of the foremost least unlikely candidate is reset to unity. Other more minor differences between the dashed and solid blue curves due to particular features of the HAYSTAC phase 1 dataset not representative of typical haloscope data are discussed in Appendix \ref{app:features}.

The figure of merit of an axion haloscope is the rate $R$ at which it can scan a given frequency window to a specified sensitivity  \cite{malnou2019squeezed}. The appropriate comparison to make between two analysis frameworks is therefore the relative amount of time $\tau_\text{rel}$ taken to perform a given scan. The dashed, gray line in Fig.\,\ref{fig:HAYSTAC data}b represents the time that would have been required to scan any coupling $g_\gamma$ to the standard $90\%$ exclusion ($10\%$ prior update) level using the FT (BT2) framework. The line derives from the $\tau_\text{rel} \propto g_\gamma^{-4}$ relation \cite{malnou2019squeezed} of scan times to couplings, and is normalized to the run-time of HAYSTAC phase 1. The inset of Fig.\,\ref{fig:HAYSTAC data}b shows all three curves in the vicinity of this line. Had the BPM framework been in place to recommend further scanning at $\nu_\text{LU}^\star$, its $\mathcal{U} = 10\%$ aggregate prior update would have taken $\approx36\%$ longer to achieve with a threshold analysis. In the context of detection efforts that take years to cover unsatisfactorily little parameter space, this is a significant enhancement available at no hardware or operational expense across a broad array of platforms.

\section{CONCLUSION} \label{sec:conclusion}

Haloscopes belong to a class of experiments that are statistics-limited. As such, optimizing the extraction of the information content of their data is a high priority. Analyzing haloscope data with threshold frameworks, however, discards information pertinent to the presence or absence of an axion. The BPM analysis framework straightforwardly applies Bayes' theorem to incorporate the relevant information content of haloscope power spectra into a posterior probability of an axion existing. Taken in ratio with the prior probability of that axion, the updated probability is seen to be completely equivalent to the standard exclusion quoted in the literature. By applying the BPM framework to the phase 1 data of the HAYSTAC experiment, we show exclusion improved commensurate with what under typical experimental conditions would translate to a $36\%$ improvement in scan time --- within the sensitivity error bounds for the HAYSTAC phase 1 dataset, though independent of the contributing sources of error \cite{brubaker2017firstA}. Additionally, the BPM framework better spotlights those regions of parameter space where the probability of an axion increases. For HAYSTAC's data, a single such frequency worthy of (but not demanding) further attention stands out. 

The BPM framework may be straightforwardly applied to future data from ADMX \cite{du2018search}, HAYSTAC \cite{zhong2018results}, and other haloscopes and similar dark matter searches \cite{lee2020axion, alesini2019galactic, kahn2016broadband,ouellet2018first, foster2018revealing, silvafeaver2017design, chaudhuri2018fundamental, mcallister2017organ, caldwell2017madmax, rybka2015search, roussy2020experimental}. We expect that BPM will probe deeper couplings at no experimental cost while permitting experimentalists to operate nimbly, without the more strict constraints of preordained scan protocols. This latter advantage is especially important for narrowband, tunable searches, in which the question of how to optimize a scan protocol by responding in real-time to acquired data is of paramount importance. Enhancements gained through operational efficiency will readily compound with those intrinsically available from BPM's informational advantage over threshold-based analyses.

\section*{Acknowledgements} 

We thank Tanya Roussy, Joshua Foster, and Lucas Sletten for helpful discussions. This work was supported by the National Science Foundation, under grants PHY-1607223 and PHY-1734006, and by the Heising-Simons Foundation under grant 2014-183.

\appendix

\section{BAYESIAN THRESHOLD FRAMEWORKS} \label{app:BT}
One of the central claims of this article is that in a haloscope analysis an apples-to-apples comparison may be made between the widely quoted frequentist exclusion $E$ of Sec.\,\ref{sec:FT} and the Bayesian prior update $U_i$ of Sec.\,\ref{sec:BPM}. In general, these two are in fact not equivalent. The equivalence holds only between FT analyses with low false-positive rates and Bayesian analyses using low priors. In this appendix, we consider two Bayesian threshold analyses. The first framework, BT1, is more informative than the second, BT2. In the dual limit of infinitesimal priors/low false-positive rates, BT2 reveals the Bayesian prior update to be completely equivalent to the frequentist exclusion. Together, BT1 and BT2 motivate the BPM framework as a more informative, less operationally constrained limit of haloscope analyses.

The first of the threshold-based Bayesian analyses we consider, BT1, updates its priors after each successive scan. We are interested in the conditional probability that there is an axion in the $i^\text{th}$ bin, given that the measured power in the $j^\text{th}$ scan either did ($x_i^{(j)} \geq x_T^{(j)}$) or did not ($x_i^{(j)} < x_T^{(j)}$) exceed threshold. For simplicity of notation, we will drop all frequency indices $i$ for the remainder of this and other appendices where not necessary, with it understood that results can ultimately be aggregated via Eq.\,\eqref{agg update}. We denote the event of a measured power exceeding (not exceeding) threshold as a binary detector going ``click" (``no-click"), and we denote the event that the measured power excess came from the axion distribution (no-axion distribution) as $\mathcal{A}$ ($\mathcal{N}$). The posterior probability of an axion given a click is 
\begin{equation} \label{P(a, click)}
    P(\mathcal{A}|\text{click}) = \frac{P(\text{click}|\mathcal{A})P(\mathcal{A})}{P(\text{click}|\mathcal{A})P(\mathcal{A}) + P(\text{click}|\mathcal{N})P(\mathcal{N})},
\end{equation}
where the denominator is equivalent to $P(\text{click})$. Identifying $P(\text{click}|\mathcal{A})$ as the single-scan true-positive rate $1-f_n^{(j)}$ (Fig.\,\ref{fig:2Gaus}, light blue) and $P(\text{click}|\mathcal{N})$ as the single-scan false-positive rate  $f_p^{(j)}$ (Fig.\,\ref{fig:2Gaus}, dark green), abbreviating the prior (posterior) probability that there is an axion in the bin as $P_a^{(0)} = P(\mathcal{A})$ [$P_{a,\text{cl}}^{(1)} = P(\mathcal{A}|\text{click})$], and noting that $P(\mathcal{N}) = 1 - P_a^{(0)}$, Eq.\,\eqref{P(a, click)} simplifies to 
\begin{equation}\label{BT update cl}
    u^{(1)}_\text{cl} = \frac{P_{a,\text{cl}}^{(1)}}{P_a^{(0)}} \approx \frac{1-f_n^{(1)}}{f_p^{(1)}} \approx \frac{1}{f_p^{(1)}}.
\end{equation}
The first approximation holds in the appropriate limit of infinitesimal priors, here $P_a^{(0)} \ll f_p^{(1)}$, and the second holds in the limit of low single-scan false-negative rates, $f_n^{(1)} \ll 1$. 

In the event of a no-click, the posterior probability $P_{a,\text{no}}^{(1)} = P(\mathcal{A}|\text{no-click})$ is given by an expression analogous to Eq.\,\eqref{P(a, click)} containing the true- ($1-f_p^{(j)}$; Fig.\,\ref{fig:2Gaus}, light green) and false- ($f_n^{(j)}$; Fig.\,\ref{fig:2Gaus}, dark blue) negative rates. The expression simplifies as 
\begin{equation}\label{BT update no}
     u^{(1)}_\text{no} = \frac{P_{a,\text{no}}^{(1)}}{P_a^{(0)}} \approx \frac{f_n^{(1)}}{1-f_p^{(1)}} \approx f_n^{(1)}, 
\end{equation}
where the first approximation holds for low priors $P_a^{(0)} \ll 1$, and the second holds for low single-scan false-positive rates, $f_p^{(1)} \ll 1$. Whereas single-scan false-negative rates in haloscope searches are typically close to $5\%$, false-positive rates are often kept much lower, to minimize the need for time-expensive rescans. 

Applied to a sequence of $M$ scans, the BT1 analysis will be seen to provide a crucial operational freedom, shared by the BPM framework, to the experimentalist. Since the ordering of multiple updates is inconsequential, a series of $M$ identical ($f_n^{(j)} = f_{n0}$, $f_p^{(j)} = f_{p0}$) scans has $M+1$ possible outcomes, corresponding to observing $c \in \{0, 1, \dots, M\}$ clicks and $M-c$ no-clicks. Using the simplest forms of Eqs.\,\eqref{BT update cl} and \eqref{BT update no}, the final prior update, obtained from Eq.\,\eqref{BPM update tot}, is 
\begin{equation} \label{BT1 U}
U^\text{BT1}_c \approx \frac{(f_{n0})^{M-c}}{(f_{p0})^c}.
\end{equation}
The prior is upgraded by $1/f_{p0}$ for every positive result, and downgraded by $f_{n0}$ for every negative result. 

In the BT1 framework, the threshold and false-negative rate for each scan must be set in advance, but the number of scans $M$ need not be. This crucial operational freedom (which is shared by the BPM framework) is a consequence of the fact that the expected value of the prior update before performing a given scan is unity, and so the decision to perform another scan or not can be made on the fly, without biasing the outcome. In contrast, in the FT framework of Sec.\,\ref{sec:FT}, adding another scan mid-way through a multiscan protocol and recalculating the false-negative rate for the entire protocol, the experimentalist can manipulate the expected value of the exclusion. This freedom has nothing to do with the use of frequentist versus Bayesian inference. We will see that Bayesian frameworks can be equally restrictive. 

\begin{figure}[!h] 
	\centering
	\includegraphics[scale=0.46]{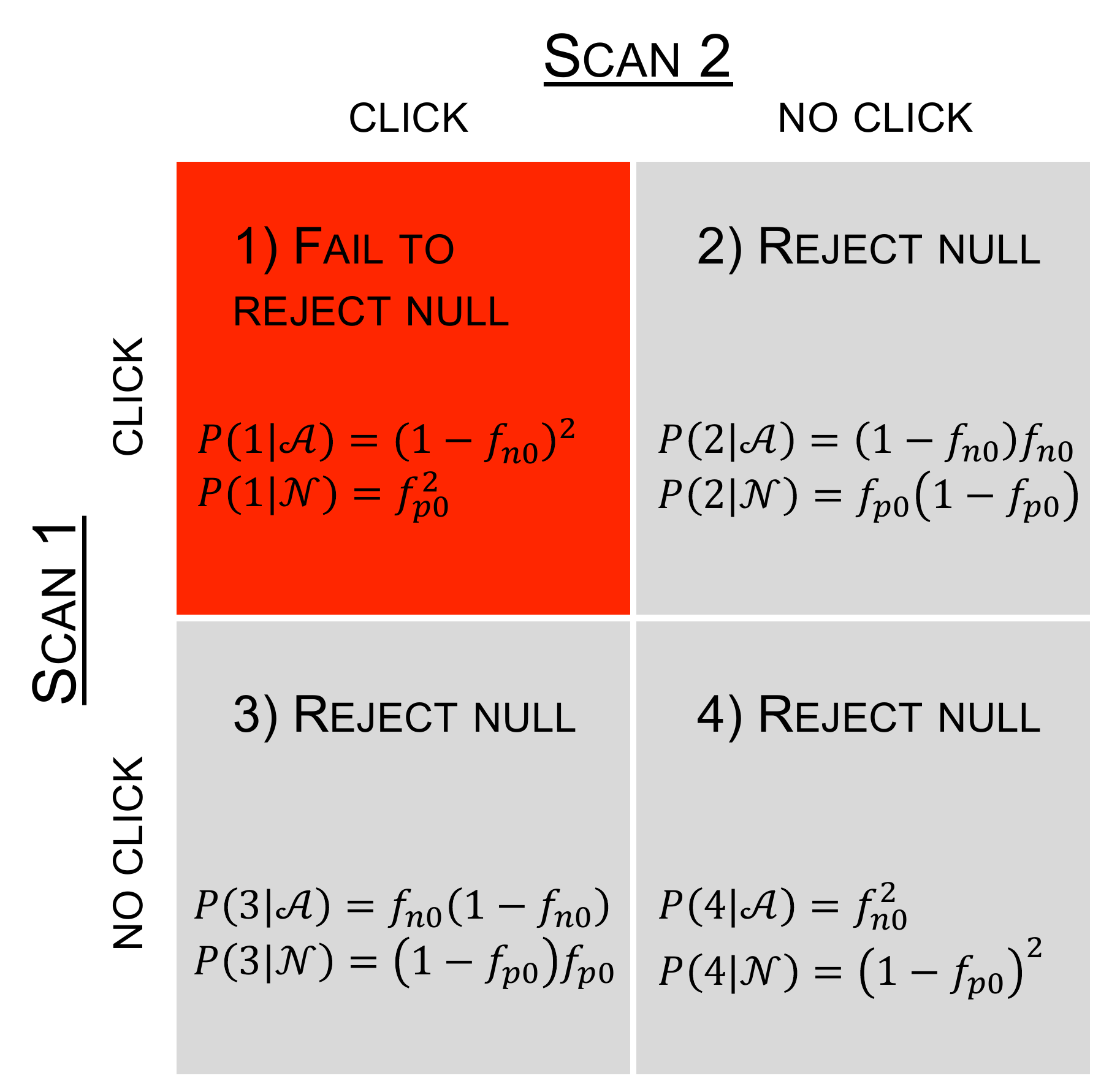}
	\caption{The four possible outcomes of an $M=2$-identical-scan threshold protocol. The conditional probabilities of each outcome $(1)$--$(4)$ subject to the axion ($\mathcal{A}$) and no-axion ($\mathcal{N}$) hypotheses are given in terms of true and false-positive and -negative rates (Fig.\,\ref{fig:2Gaus}). In both the FT framework of Sec.\,\ref{sec:FT} and the BT2 framework of  Appendix \ref{app:BT}, outcomes $(2)$--$(4)$ (gray) are all treated identically as the negative result: in the FT (BT2) framework, the null is rejected (the prior is reduced). In this case, the exclusion $E = 1-F_n$ is trivially equivalent to the prior update $U^\text{(BT2)}_\text{neg} \approx F_n$ in the dual limit of low false-positive rates and priors. Only region $(1)$ (red) fails to reject the null (FT), or increases the prior (BT2).}
	\label{fig:2Scan}
\end{figure}

Whereas the BT1 framework offers $M+1$ possible outcomes for $M$ scans and provides interscan operational freedom to the experimentalist, the second BT framework (BT2) that we consider is designed to be uninformative and restrictive by comparison; its outcomes, tellingly, will map precisely onto those of the FT framework of Sec.\,\ref{sec:FT}. The standard FT analysis permits two possible results at each frequency, illustrated for a simple case in Fig.\,\ref{fig:2Scan}: a rejection of the null (negative result) or a failure to reject it (positive result). For an $M$-scan protocol, the negative result is defined as the case where at least one scan no-clicks; the positive result occurs only when all $M$ scans click. The multiscan false-negative rate, $F_n$ from Eq.\,\eqref{tot F neg}, sets the exclusion via Eq.\,\eqref{excl}. False-positive rates do not enter directly into the FT framework's reported exclusion.\footnote{More completely, false-positive rates only enter indirectly, as the global false-positive rate --- the chance of at least one bin achieving a positive result (i.e. a click on all $M$ scans) --- must be set sufficiently low to account for the look-elsewhere effect. Choosing a higher false-positive rate will achieve a deeper exclusion if no positive results are recorded, but will increase the probability of such an event.} 

The BT2 framework applies Bayes' theorem, Eq.\,\eqref{Bayes}, to this binary result landscape, obtaining a posterior either for a positive result of $P_{a,\text{pos}}^{(1)}$ or for a negative result of $P_{a,\text{neg}}^{(1)}$. For the positive case,
\begin{equation} \label{P(a, pos)}
    P(\mathcal{A}|\text{pos}) = \frac{P(\text{pos}|\mathcal{A})P(\mathcal{A})}{P(\text{pos}|\mathcal{A})P(\mathcal{A}) + P(\text{pos}|\mathcal{N})P(\mathcal{N})}
\end{equation}
yields
\begin{equation} \label{U BT2 pos}
    U^{\text{BT2}}_\text{pos} = \frac{P_{a,\text{pos}}^{(1)}}{P_a^{(0)}} \approx \frac{1-F_n}{F_p}\approx \frac{1}{F_p},
\end{equation}
where the total false-negative rate $F_n$ is that of Eq.\,\eqref{tot F neg} and 
\begin{equation} \label{tot F pos}
    F_p = \prod_{j=1}^M f_p^{(j)}
\end{equation}
is the total false-positive rate for the $M$ scans. The first (second) approximation in Eq.\,\eqref{U BT2 pos} is valid for infinitesimal priors, $P_a^{(0)} \ll F_p$ (low total false-negative rate, $F_n \ll 1$). Note that BT2's positive prior update agrees in the case of identical scans (the general cases agree as well) precisely with the $c = M$ case of Eq.\,\eqref{BT1 U} for BT1. The agreement is due to the fact that they derive from the same event --- a series of $M$ clicks. Since false-positive rates are very small, $U_\text{pos}^\text{BT2}$ is a very large number, corresponding to just how unlikely a positive result would be to observe, without an axion present. 

The negative result yields an update that will agree precisely with the standard FT exclusion $E$. From the equation for $P(\mathcal{A}|\text{neg})$ analogous to Eq.\,\eqref{P(a, pos)}, we obtain:
\begin{equation} \label{U BT2 neg}
    U^{\text{BT2}}_\text{neg} = \frac{P_{a,\text{neg}}^{(1)}}{P_a^{(0)}} \approx \frac{F_n}{1-F_p}\approx F_n,
\end{equation}
where the two approximations assume low priors $P_a^{(0)} \ll 1$ and low $M$-scan false-positive rates $F_p \ll 1$. With these conditions met, the BT2 negative prior update is the complement of the frequentist exclusion, $U_\text{neg}^\text{BT2} \approx 1 - E$. In the limit relevant for haloscope searches, the frequentist exclusion is therefore equivalent to the Bayesian prior update, an equivalence used in the context of other searches for new physics \cite{lista2017statistical}. Whether this update is obtained through one of the BT frameworks or the BPM framework is immaterial. 

Finally, a pair of differences between the two BT frameworks discussed in this appendix hints at two important operational advantages of the BPM framework. The BT1 framework, by allowing $M+1$ possible results with differing outcomes (or up to $2^M$ for scans with nonidentical false-negative and -positive rates), has an informational advantage over the BT2 framework. The BT2 framework deliberately blinds the experimentalist to which scan(s) no-clicked, combining $M$ of the $M+1$ possible outcomes into a single negative result. For an experimentalist intent on discovering or excluding the axion, there can be no benefit to discarding this information. Taking the next logical step, the power-measured information from each scan need not be discarded. The BPM framework, unlike the BT frameworks, uses this information. 

\begin{table}[]
    \centering
    \begin{tabular}{|l|l|l|l|} 
    \hline
    & \textbf{BT2 (FT)}& \textbf{BT1} & \textbf{BPM}\\ \hline
    possible outcomes per bin & $2$ & $2^M$ & $\infty$ \\ \hline
    interscan freedom & no & yes & yes \\ \hline
    intrascan freedom & no & no & yes \\ \hline
    \end{tabular} 
    \caption{Comparison of the Bayesian threshold frameworks (BT1 and BT2) of Appendix \ref{app:BT} with the Bayesian power measured (BPM) framework of Sec.\,\ref{sec:BPM} for a protocol with $M$ possibly nonidentical scans performed. More informative frameworks map the continuum of possible measurements onto a larger number of reported outcomes. The more informative frameworks also permit the experimentalist greater freedom to alter the scan protocol without biasing the outcome. Interscan changes to upcoming scans are allowed within all but the BT2 framework (equivalently, the frequentist threshold (FT) framework of Sec.\,\ref{sec:FT}), while the BPM framework even permits intrascan adjustments as information compiles.} \label{tab:Bayes compare}
\end{table}

The BT1 framework also proves operationally superior to BT2. Whereas the BT2 framework locks in a commitment to $M$ scans, if necessary, BT1 allows the experimentalist to reinsert himself into the decision-making \textit{between} scans without biasing the outcome. This interscan freedom is improved to intrascan freedom for BPM, as indicated in Table \ref{tab:Bayes compare}, which summarizes the informativeness and operational constraints imposed by each framework. Under BPM, the expected prior update for the next iota of power measured always being unity protects the experimentalist from inserting bias. To the degree that the data processing allows it, probability can be tracked in real time and used to inform scan-protocol decisions on the fly. This capability provides an opportunity for optimizing a BPM haloscope search algorithm, a promising direction for future analysis beyond the scope of this article.

\section{OUTCOME DISTRIBUTIONS FOR BAYESIAN ANALYSES}
\label{app:expected outcomes}

The reanalysis of the HAYSTAC phase 1 dataset in Sec.\,\ref{sec:HAYSTAC} indicates that the BPM framework achieves superior exclusion to thresholding in the case of one real dataset. In this appendix, we demonstrate that this result is typical by treating the prior updates in the BPM framework and the  BT1 and BT2 frameworks discussed in Appendix \ref{app:BT} as random variables. We make an apples-to-apples comparison between the aggregate prior update probability distributions obtained from an $M=2$-identical scan protocol in these three frameworks. In particular, while the mean aggregate prior update $\braket{\mathcal{U}} = 1$ for any unbiased analysis, we will see that the \textit{median} aggregate prior update in the BPM framework is a factor-of-two smaller than in either threshold framework. From this reduction in the median aggregate prior update we predict a typical scan rate enhancement of $30\%$ for BPM relative to thresholding, consistent with the enhancement observed in the HAYSTAC phase 1 dataset. En route to this final result, we obtain analytic expressions for the total prior update probability distributions in each framework. Comparing the three frameworks at this level likewise elucidates the difference between the BPM and threshold exclusion lines plotted in Fig.\,\ref{fig:HAYSTAC data}a.

Throughout this appendix, we assume an ideal, axionless haloscope dataset: the grand spectrum excesses are independent, identically distributed random variables $x_i^{(j)} \sim N(0,1)$. We compare prior update probability distributions for an $M=2$-identical scan protocol in which an initial scan is performed, followed by a rescan in each bin whose measured power excess exceeds a predetermined threshold.\footnote{As discussed in Appendix \ref{app:BT}, BPM provides freedoms to deviate from such a rigid protocol, and can therefore outperform the projections here.} For definiteness, we assume scan parameters similar to those used in HAYSTAC phase 1. For the BT1 and BT2 frameworks, we will first derive analytic expressions for the prior updates in a more general $M$-identical scan protocol, and then specify to $M=2$.

Starting with the least informative BT2 framework, the total prior update in a single bin has probability mass function (PMF):
\begin{align} 
\begin{split}\label{p U BT2}
    P(U^\text{BT2} &= U^\text{BT2}_\text{pos}) = (f_{p0})^M \\
    P(U^\text{BT2} &= U^\text{BT2}_\text{neg}) = 1 - (f_{p0})^M.
\end{split}
\end{align}
In any scan of $N$ independent frequency bins, $k = 0,\dots,N$ of those will realize a positive result, where $k$ is binomial distributed $k\sim B(N, (f_{p0})^M)$. The PMF for the aggregate prior update is then
\begin{align}
\begin{split}\label{p agg U BT2}
    P\left(\mathcal{U}^\text{BT2} = \frac{U_\text{pos}^\text{BT2} - U_\text{neg}^\text{BT2}}{N} k + U_\text{neg}^\text{BT2} \right) = \\
    {N \choose k}[(f_{p0})^M]^k[1-(f_{p0})^M]^{N-k}.
\end{split}
\end{align}

Next we consider the BT1 framework, whose total prior update $U_c^\text{BT1}$, Eq.\,\eqref{BT update cl}, is written in terms of the single-scan updates $u_\text{cl}$ and $u_\text{no}$ for a click, Eq.\,\eqref{BT update cl}, and for a no-click, Eq.\,\eqref{BT update no}, respectively. Each click occurs with probability $f_{p0}$, hence the probability of obtaining exactly $c = \{0,1,\dots,M\}$ clicks is 
\begin{equation} \label{pc}
    P_c = (f_{p0})^c (1-f_{p0})^{1-\delta_{Mc}},
\end{equation}
where $\delta_{ab}$ is the Kronecker delta function. The total prior update PDF is therefore 
\begin{equation}\label{p tot U BT1}
    P\left(U^\text{BT1} = [u_\text{cl}]^c [u_\text{no}]^{1-\delta_{Mc}} \right) = P_c.
\end{equation}
For the $M$-identical-scan protocol over $N$ independent bins, the aggregate prior update (Eq.\,\eqref{agg update}) is parametrized by the numbers $n_c$ of bins that click $c$ times, which are multinomial distributed with probabilities $P_c$ given by Eq.\,\eqref{pc}. The aggregate prior update is 
\begin{multline}\label{p agg U BT1}
P\left(\mathcal{U}^\text{BT1} = \frac{1}{N} \sum_{c=0}^M n_c \frac{(u_\text{cl})^c}{(u_\text{no})^{\delta_{Mc}-1}}\right) = \\ 
\frac{N!}{\prod_{c^\prime=0}^M n_{c^\prime}!} \prod_{c^{\prime\prime} = 0}^M (P_{c^{\prime\prime}})^{n_{c^{\prime\prime}}}
\end{multline}
if $\sum_{c=0}^M n_c = N$, and $0$ otherwise.

\begin{figure*}[t] 
	\centering
	\includegraphics[scale=1.00]{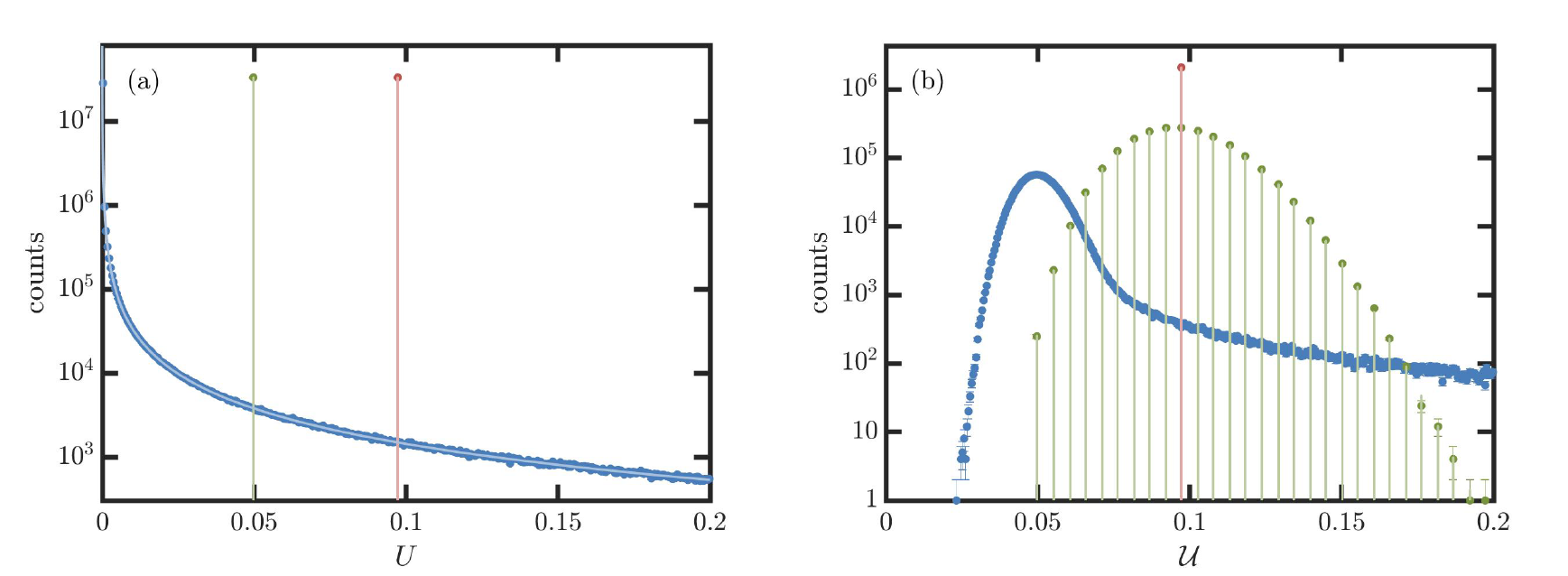} 
	\caption{Outcome probability distributions for prior updates for an axion that would produce a mean excess $\mu_a=5.1$, assuming there is in reality no axion present and using a standard $M=2$-identical-scan protocol with scan parameters similar to those of HAYSTAC: threshold $x_T = 3.455$ ($f_{n0} = 5\%$, $f_{p0} = 0.03\%$). (a) Histograms of simulated total prior updates $U$, Eq.\,\eqref{BPM update tot}, for the BPM (blue dots), BT1 (green dots), and BT2 (red dots) frameworks. $2^{25}$ independent bins were simulated, and error bars estimated from binomial statistics are smaller than the data points. The simulated probability distributions are in excellent agreement with the analytic predictions of Eqs.\,\eqref{f U BPM full}, \eqref{p tot U BT1}, and \eqref{p U BT2} (blue, green, and red lines). The total updates described in (a) are combined via Eq.\,\eqref{agg update} to form the aggregate prior updates $\mathcal{U}$ described in (b). The aggregate prior updates are numerically generated over $2^{21}$ independent trials of $N=2^{15}$ independent bins for each of the three analysis frameworks. For the frameworks for which we derive analytic predictions (green and red lines for BT1, Eq.\,\eqref{p agg U BT1}, and BT2, Eq.\,\eqref{p agg U BT2}, respectively), the agreement with the simulations within binomial error bars confirms our expectations. For all three frameworks, the (relatively few) bins which fall outside of the plotted windows do indeed balance the scales so as to produce the unbiased expectation values $\braket{U} = \braket{\mathcal{U}} = 1$ (Appendix \ref{app:corr}). The BPM framework displays by far the lowest median outcomes in all cases, indicating that it will typically achieve superior exclusion when no axion is present.}
	\label{fig:Outcome Dists}
\end{figure*}

For the BPM framework, we restrict ourselves to $M=2$ identical scans, where the second occurs conditionally on the first exceeding threshold, and derive only the expression for the total prior update. To derive the PDF for the two-scan BPM prior update, we consider two mutually exclusive cases. 

First, with probability $1-f_{p0}$, the initial scan does not exceed threshold ($x<x_T$; a true-negative). In that case, the prior update PDF is 
\begin{align} 
\begin{split}\label{f u true neg}
    f_{u,\text{tn}}(u_\text{tn}) &= f_x(x)\left(\frac{du}{dx}\right)^{-1}\\
    &= \begin{cases} 
    (1-f_{p0})^{-1}f_u(u) & 0\leq u_\text{tn}<u_T\\
    0 & \text{otherwise},
    \end{cases}
\end{split}
\end{align}
where 
\begin{equation} \label{f u}
    f_u(u) =
    \begin{cases}\left(\frac{\exp[-\mu_a^2/8]}{\mu_a\sqrt{2\pi}}\right)  \left(\frac{\exp[-(\log u)^2/2\mu_a^2]}{u^{3/2}}\right) & 0 \leq u\\ 
    0 & u<0
    \end{cases}
\end{equation}
and $u_T = \exp[-\mu_a^2/2 + \mu_a x_T]$ is the single-scan prior update achieved at threshold ($x=x_T$). The second line of Eq.\,\eqref{f u true neg} follows from Eqs.\,\eqref{pdf} and \eqref{BPM update}. The normalization factor $(1-f_{p0})$ accounts for the nonunit probability of drawing from this distribution in the initial scan. 

Next we consider the case where the first scan exceeds threshold, ($x \geq x_T$; a false positive), which occurs with probability $f_{p0}$. The total prior update for the initial false-positive is the product of two single-scan prior updates $U_\text{fp} = u_\text{fp} u$. The PDF for the first scan prior update $u_\text{fp}$ is
\begin{equation} 
\label{f u false pos}
    f_{u,\text{fp}}(u_\text{fp}) = 
    \begin{cases} 
    (f_{p0})^{-1}f_u(u_\text{fp}) & u_T\leq u_\text{fp}\\
    0 & u_\text{fp} < u_T,
    \end{cases}
\end{equation}
and that for the second scan prior update $u$, which is not restricted to being above or below threshold, is given by Eq.\,\eqref{f u}.

The PDF for $U_\text{fp}$ is then
\begin{align}
\begin{split} \label{f U false pos}
    &f_{U,\text{fp}}(U_\text{fp}) = \int_{-\infty}^\infty f_u(u) f_{u,\text{fp}}\left( \frac{U_\text{fp}}{u}\right) \frac{du}{|u|} \\
    &= \begin{cases}
    \frac{\exp\left[-\left(\mu_a/2\right)^2 - \left(\log\left[U_\text{fp}/2\mu_a \right]\right)^2\right]}{f_{p0}\mu_a U_\text{fp}^{3/2} 4 \sqrt{\pi}}\text{erfc}\left(\frac{\log\left[u_T^2/U_\text{fp}\right]}{2\mu_a}\right) & 0 \leq U\\
    0 & U<0, 
    \end{cases}
\end{split}
\end{align}
where $\text{erfc}$ is the complementary error function. 

The total prior update $U$ for the procedure is a probability-weighted sum of PDFs: \begin{equation}
f_U(U) = (1-f_{p0})f_{u,\text{tn}}(U) + f_{p0}f_{U,\text{fp}}(U).
\end{equation}
Written out fully,
\begin{widetext}
\begin{equation} \label{f U BPM full}
    f_U(U) = \begin{cases}
    \frac{\exp\left[-\left(\mu_a/2\right)^2 - \left(\log\left[U/2\mu_a \right]\right)^2\right]}{\mu_a U^{3/2} 4 \sqrt{\pi}}\text{erfc}\left(\frac{\log\left[u_T^2/U\right]}{2\mu_a}\right) +  \frac{\exp[-\mu_a^2/8 -(\log U)^2/2\mu_a^2]}{\mu_a U^{3/2} \sqrt{2\pi}} & 0 \leq U < u_T \\

    \frac{\exp\left[-\left(\mu_a/2\right)^2 - \left(\log\left[U/2\mu_a \right]\right)^2\right]}{\mu_a U^{3/2} 4 \sqrt{\pi}}\text{erfc}\left(\frac{\log\left[u_T^2/U\right]}{2\mu_a}\right) & u_T \leq U\\
    
    0 & U<0
    \end{cases}
\end{equation}
\end{widetext}
is the two-identical-scan BPM total prior update PDF: that is, the probability density for obtaining a prior update $U$ in a single axionless bin using the standard initial and conditional-rescan procedure.  The total updates for larger numbers of scans and the aggregate updates for multiple bins are too cumbersome to write out in closed form, but may readily be simulated. 

Equations \eqref{f U BPM full}, \eqref{p tot U BT1}, and \eqref{p U BT2} are respectively the total prior update probability distributions for the BPM framework and the two BT frameworks discussed in Appendix \ref{app:BT}. These three equations are plotted as blue, green, and red lines, respectively, alongside simulated total prior updates for the three frameworks (blue, green, and red dots) for a two-scan protocol without an axion present in Fig.\,\ref{fig:Outcome Dists}a. 

The simulations are performed by drawing Gaussian random variables according to Eq.\,\eqref{pdf}, with $\mu=0$, $\sigma=1$, and performing a second draw as well if $x \geq x_T$. Prior updates are then applied according to Eqs.\,\eqref{BPM update}, \eqref{BPM update tot}, \eqref{BT1 U}, \eqref{U BT2 pos}, and \eqref{U BT2 neg}. The simulations match the analytic expressions, and together reveal that the vast majority of bins will receive a downward update under all three frameworks. In particular, the total prior updates for BT1 and BT2 almost always take the precise values $f_n$ and $1-(1-f_n)^2$, respectively. The price we pay for these reliably low typical updates are rare false positives with  $U \gg 1$ (far off the right side of the plot). The common $U < 1$ and rare $U \gg 1$ outcomes together ensure $\braket{U}=1$, the condition for an unbiased analysis. In contrast, moderate positive and negative prior updates occur much more frequently in the BPM framework, and $\braket{U}=1$ is enforced by a long tail of positive prior updates. The fact that the median of the BPM distribution is lower than that of either threshold distribution is noteworthy, as we will see presently in our discussion of the aggregate prior update.

Figure \ref{fig:Outcome Dists}b shows the aggregate prior updates $\mathcal{U}$ for the identical two-scan protocol when there are $2^{15}$ independent bins, not far from the actual number in HAYSTAC \cite{brubaker2017haystac}. In all cases, the total updates from all bins are averaged according to Eq.\,\eqref{agg update} to obtain the aggregate prior update. All three methods deliver an unbiased aggregate prior update with mean $\braket{\mathcal{U}}=1$, yet the BPM framework typically delivers a stronger exclusion in the absence of an axion, i.e. its \textit{median} aggregate prior update is smaller than in either threshold framework by a factor-of-two. The improvement is a consequence of the informational advantage of the BPM framework.

The factor-of-two typical improvement in aggregate exclusion illustrated for typical HAYSTAC parameters in Fig.\,\ref{fig:Outcome Dists}b yields a scan rate enhancement consistent with the $36\%$ speedup inferred from real data in Fig.\,\ref{fig:HAYSTAC data}b, absent an axion.\footnote{With an axion present, additional simulations (not shown) reveal its presence in a manner consistent with the interpretation of the global false-positive rate used in frequentist hypothesis testing.} If the time for the BT2 framework to achieve the targeted $90\%$ exclusion across the scan window (factor-of-ten reduction in aggregate probability) is $T_{90}$, the additional factor-of-two should take approximately $T_{95} = T_{90} \log_{10}(2)$, roughly $30\%$ longer.

\section{AGGREGATE EXCLUSION} \label{app:agg excl}
The aggregation of the frequency-dependant Bayesian prior updates, $U_i(g_\gamma) \rightarrow \mathcal{U}(g_\gamma)$ via Eq.\,\eqref{agg update} has a useful frequentist counterpart for the FT framework discussed in Sec.\,\ref{sec:FT}. In this appendix, we motivate and define the aggregate exclusion $\mathcal{E}(g_\gamma)$ as the analog of $\mathcal{U}(g_\gamma)$, extending the apples-to-apples comparison of FT to BPM in Appendix \ref{app:BT} to aggregated results. Additionally, we show that $\mathcal{E}(g_\gamma)$ provides a more conservative, yet more faithful assessment of the exclusion achieved by a haloscope than does the standard practice of averaging the couplings at which a given exclusion $E$ is achieved. 

The aggregate exclusion $\mathcal{E}(g_\gamma)$ answers the same question that the exclusion $E_i(g_\gamma)$, here written explicitly as dependent upon coupling $g_\gamma$ and frequency bin $i$, does, but for the entire frequency range. The exclusion $E_i(g_\gamma)$ answers the question: ``supposing an axion of coupling $g_\gamma$ occupies bin $i$, what is the probability of observing a predefined positive result there?" We therefore define the aggregate exclusion $\mathcal{E}(g_\gamma)$ as the chance of observing a predefined, aggregate positive result (defined below), assuming an axion to exist within exactly one bin within the range. In keeping with the uniform priors assumption of Appendix \ref{app:priors}, the likelihood of an axion to be at any given bin is equal for all bins, and for simplicity the powers in bins are assumed otherwise uncorrelated (Appendix \ref{app:corr}). Since it is not known which bin the axion is in, the aggregate positive result is defined to mean at least one positive result across the $N$ frequency bins.\footnote{In other words, in the case where any nonzero number of bins click, we fail to reject the hypothesis that there is an axion somewhere within the window.} In the logic of Sec.\,\ref{sec:FT}, the known sensitivities $\eta_i^{(j)}$ and predetermined power thresholds $x_T^{(j)}$ together preselect couplings $g_{\gamma,i}$ to test at a uniform, desired false-negative rate $F_n$. This logic can be partially inverted: assuming the same $\eta_i^{(j)}$ and $x_T^{(j)}$, now use a spectrally uniform coupling $g_\gamma$ to preselect frequency-dependent total false-negative rates $F_{n,i}(g_\gamma)$ at which to test. This framing allows for the testing of a uniform coupling $g_\gamma$ across the entire range to a confidence level of
\begin{align} 
\begin{split}\label{agg excl}
\mathcal{E}(g_\gamma) &= 1 - (1-F_p)^N\frac{1}{N} \sum_{i=1}^N F_{n,i}(g_\gamma) \\ &\approx \frac{1}{N} \sum_{i=1}^N E_i(g_\gamma),
\end{split}
\end{align}
where the approximation is valid when the total false-positive rate, Eq.\,\eqref{tot F pos}, satisfies $F_p \ll 1/N$. The confidence tested at, shown for the HAYSTAC phase 1 sensitivity profile by the red curve in Fig.\,\ref{fig:HAYSTAC data}b, increases monotonically with the coupling $g_\gamma$. The coupling coincident with a desired percentile (e.g. $90\%$) can be determined from the sensitivity profile $\eta_i^{(j)}$ alone, and a standard confidence test performed by comparing the measured power excesses $x_i^{(j)}$ against the predetermined thresholds $x_T^{(j)}$, with a positive result defined as a click on all $M$ scans in at least one bin. Defined as such, the aggregate exclusion $\mathcal{E}(g_\gamma)$ bears exactly the same relationship to the aggregate prior update $\mathcal{U}(g_\gamma)$ that $E_i(g_\gamma)$ bears to $U_i(g_\gamma)$, making it central to comparing analysis frameworks. 

The aggregate exclusion, by answering the question of frequentist confidence testing, proves a more natural figure of merit for FT exclusion over a given scan window than the standard practice \cite{zhong2018results} of quoting the average coupling $g_\gamma^\text{avg}$ excluded to a given confidence $E$. If the aggregate coupling $\mathcal{G}_\gamma$ is defined as that for which identical aggregate exclusion $\mathcal{E}(\mathcal{G}_\gamma) = E$ is achieved, then a comparison will reliably reveal $\mathcal{G}_\gamma > g_\gamma^\text{avg}$.\footnote{An example for a simple $N=2$-bin grand spectrum using an $M=1$-scan protocol will suffice to illustrate how it is that averaging in the space of confidence levels, Eq.\,\eqref{agg excl}, may produce inherently conservative results relative to averaging in the space of couplings excluded at. If one bin has sensitivity to test at 90\% confidence for a $g_\gamma = g_\gamma^\text{KSVZ}$ axion (using distribution mean $\mu_a = 5.1$) and a second has sensitivity to perform the same test for a $g_\gamma = 3g_\gamma^\text{KSVZ}$ axion, then the average coupling tested at $90\%$ is $2g_\gamma^\text{KSVZ}$. However, the bin with KSVZ-sensitivity is tested at $100\%$ confidence ($F_n = 0$) for the $2$-KSVZ axion, whereas the bin with $3$-KSVZ-sensitivity is tested at only $6\%$ confidence ($F_n = 94\%$) for the $2$-KSVZ axion. Therefore the hypothesis that a $2$-KSVZ axion lies in either bin is tested not at $90\%$ but at a far weaker confidence of $53\%$.} The standard metric, $g_\gamma^\text{avg}$, is not somehow achieving deeper sensitivity, it is averaging in a space inconvenient to the natural logic of frequentism, and biased relative to it. More directly, an experiment that achieves $90\%$ exclusion at a frequency-averaged coupling of $g_\gamma^\text{KSVZ}$ has \textit{not} rejected the hypothesis that a KSVZ axion lies within its scan window with $90\%$ confidence.

\section{CORRELATIONS} \label{app:corr}
In this appendix, we argue that correlations between grand spectrum powers at different frequencies do not bias the BPM framework's outputs, and are furthermore desirable. The main text and other appendices treat the grand spectrum power excesses as uncorrelated for simplicity. In practice, this assumption is violated, as grand spectrum bin spacing are intentionally made spectrally smaller than the axion linewidth and the frequency window of the digital high-pass SG filter applied to the data. The effect of estimating power excesses based on the axion lineshape (Step \ref{step: reb2grand} in Appendix \ref{app:combine}) is to positively correlate nearby bins, whereas the effect of the SG filtering (Step \ref{step: raw2proc}) is to negatively correlate bins over a somewhat longer frequency scale \cite{brubaker2017haystac}. These are the only two processing steps that will correlate adjacent bins.\footnote{Systematic effects in the measurement will also correlate grand spectrum power excesses. The identification and removal of nonaxion power excesses and the SG filtering exist largely to undo such experimentally-induced correlations. We do not consider the effect of such correlations on the grounds that these processing steps appear largely successful \cite{brubaker2017haystac}.} The data processing introduces no interscan correlations, as it handles separate scans independently. The single-scan prior updates $u_i^{(j)}(g_\gamma)$ can therefore be treated as unconditionally unbiased ($\braket{u_i^{(j)}} = 1$), scan-independent, frequency-dependent random variables.  

Given these properties of the single-scan prior updates $u_i^{(j)}(g_\gamma)$, the result that the aggregate prior update $\mathcal{U}(g_\gamma)$ is unbiased then follows from the fact that aggregation entails multiplying across scans and summing across bins. The total prior updates $U_i(g_\gamma)$ are defined in Eq.\,\eqref{BPM update tot} as the product of grand spectrum prior updates from $M_i$ scans. Using the fact that the expectation value of the product of \textit{independent} random variables is the product of the expectation values, here
\begin{equation}\label{expect total update}
    \braket{U_i(g_\gamma)} = \left\langle\prod_{j=1}^M u_i^{(j)}(g_\gamma)\right\rangle = \prod_{j=1}^M\braket{u_i^{(j)}(g_\gamma)} = 1,
\end{equation}
the total prior updates are seen to be unbiased, i.e. have unit expectation. Aggregation then occurs through summation of the posteriors and priors via Eq.\,\eqref{agg update}. The expectation value of the sum of \textit{dependent} random variables is the sum of the expectation values, here 
\begin{equation}\label{expect agg update}
    \braket{\mathcal{U}(g_\gamma)} = \left\langle\sum_{i=1}^NU_i(g_\gamma)\right\rangle = \sum_{i=1}^N\braket{U_i(g_\gamma)} = 1,
\end{equation}
where the last equality uses the result of Eq.\,\eqref{expect total update} to demonstrate that $\mathcal{U}(g_\gamma)$ is also unbiased. As we relax assumptions of uniform or even low priors, the relevant updates, treated as random variables, can be shown to have unit expectation independent of grand spectrum correlations, but the formulae used to compute and combine the updates become less elegant. 

The primary effect of grand spectrum correlations on the prior updates is not to bias the analysis, but only to alter the higher moments of the distribution that aggregate prior updates are drawn from. In particular, the number $N_I$ of \textit{independent} grand spectrum bins is smaller than the number $N$ of grand spectrum excesses calculated. This increases the spread of the distribution that aggregate prior updates will be drawn from for a given haloscope scan window. However, since the dominant correlations are produced by the axion lineshape (Step \ref{step: reb2grand}, Appendix \ref{app:combine}), the number of independent bins has as a rough lower bound the number of axion linewidths  $\Delta_a$ in the scan window $\Delta_W$, and as an upper bound the number of bins used: $\Delta_W/\Delta_a \lsim N_I \leq N$. The fact that $N_I \leq N$ broadens the probability distributions of the aggregate outcomes (Appendix \ref{app:expected outcomes}), but does not otherwise impact the analysis.  

To see that correlations are indeed desirable, note that their absence would imply a grand spectrum bin spacing larger than the axion linewidth. As discussed in Ref.\,\cite{brubaker2017haystac}, such a coarse bin spacing would effectively leave unprobed a large fraction of the scan window. Since the correlations produce no deleterious effects upon the BPM framework, there is no reason not to oversample the space of possible axion masses, and the correlations should be considered a feature, not a bug.

\section{PRACTICAL DEPARTURES FROM GAUSSIANITY}
\label{app:non-Gauss}
The formula for updating priors using the BPM framework, Eq.\,\eqref{BPM update}, assumes that the axion and no-axion distributions are Gaussian, Eq.\,\eqref{pdf}. In particular, measured distributions often rise in their extreme tails over what is ideally expected \cite{bevington2003data}. In the case of haloscopic axion detection, the presence of spurious rf tones increases the probability density of the high-power tail of the PDFs of Fig.\,\ref{fig:2Gaus}. In this appendix, we argue that the interpretation of the exclusion achieved with the BPM framework is robust to the resultant bias towards discovery. Conversely, there is little to be gained from Bayesian inference in the context of manually interrogating a persistent signal with the goal of reporting discovery. Efficiency is a far less important consideration in this context, and test statistics such as those of Refs.\,\cite{beaujean2018bump, foster2018revealing}, in conjunction with manual interrogation (Footnote \ref{FN: manual}) are well suited to the task of claiming discovery. 

Sources of rf noise in the haloscope's physical environment can couple into the receiver chain. Real haloscopes go to significant effort to discriminate such artificial excesses from real axion signals, but excesses sufficiently similar in spectral profile to an axion necessarily contribute to the measured power distributions, leading to an effective increase in the high power probability density of both the axion and no-axion distribution PDFs. 

The effect of increased probability density in the high-power tail in either distribution in Fig.\,\ref{fig:2Gaus} is to bias the BPM analysis towards discovery of an axion. In any one frequency bin, this effect is very unlikely because it only affects the distribution tails. Indeed, for existing haloscopes, the measured power excesses seem very well approximated out to several standard deviations by Gaussian distributions \cite{asztalos2001large, brubaker2017haystac}, indicating that the consequences of non-Gaussian features will be modest. However, with tens or hundreds of thousands of independent frequency bins measured, there may be a small augmentation of the BPM framework's aggregate prior update, Eq.\,\eqref{agg update}.

In the context of reporting exclusion over axion parameter space --- the only context in which data from haloscopic searches has appeared to date --- a bias towards discovery is conservative, and hence acceptable. In the context of a FT framework, the effect of a discovery bias is not to change the reported exclusion, but rather to increase expected rescan time and the chance of a false-positive. Since, as shown in Fig.\,\ref{fig:HAYSTAC data}b, the BPM framework still achieves a $36\%$ scan rate enhancement relative to thresholding for the HAYSTAC phase 1 dataset, the conservative bias is manifestly not large enough (at least for this dataset) to overwhelm the improved exclusion over the whole scan range.

In the context of reporting the discovery of axionic dark matter, conversely, a bias towards discovery would appear to be harmful. However, the nature of axion detection guards against adverse outcomes in practice. A power excess presenting as an axion, once identified, is fairly straightforwardly shown to persist or not by manual interrogation. The role of the statistical analysis frameworks discussed in this article is not to firmly identify the axion as such once a persistent signal is found, but rather to indicate with the highest fidelity where such a signal is likely to lie. While a BPM analysis may bias the experimentalist towards believing that a few nonaxionic excesses might be axionic, the real certification of an axion signal would look like that for other discoveries in fundamental physics: a series of careful measurements precluding beyond doubt that no other known physical phenomena could account for the signal. In other words, the asymmetry between the criteria for exclusion and discovery ensure that BPM's discovery-bias would never lead to accidental discovery.

\section{CHOICE OF PRIORS} \label{app:priors}
For any choice of prior probability $P_{a,i}^{(0)}$ in axionic dark matter, Bayes' theorem can be applied and updates $U_i$ calculated and reported. However, the utility of the BPM framework rests on the fact that its prior updates $U_i$ can be treated as independent of the priors chosen for the axion and competing hypotheses, i.e. as reasonably approximated by Bayes factors. Additionally, as argued in Appendix \ref{app:BT}, it is highly convenient that in the appropriate limit the prior updates of the BT2 framework, itself directly comparable to the BPM framework, become the conditional probabilities quoted in frequentist exclusion. These conveniences hold only so long as one's priors are infinitesimal but nonzero.\footnote{The fact that the prior must be nonzero is justified by Cromwell's rule, which states that priors of exactly 0 or 1 should be avoided generally. More directly, there would be no sense in performing a haloscope search if the chance of success was truly zero.} In this appendix, we justify the assumption of infinitesimal priors, providing a toy estimate of a prior for illustrative purposes. Additionally, we discuss how the prior update can be used to directly inform scan strategy in the presence of nonaxionic rf power excesses. 

The constraint that the prior be infinitesimal is only used in simplifying the denominator of Bayes' theorem, Eq.\,\eqref{Bayes}, according to the law of total probability:
\begin{align}
\begin{split}\label{law tot prob}
    P\left(x_i^{(j)}\right) &=
    \int_0^\infty P\left(x_i^{(j)}|\mathcal{A}_i(g_\gamma)\right)P\left(\mathcal{A}_i(g_\gamma)\right)dg_\gamma \\&+ P\left(x_i^{(j)}|\mathcal{N}_i\right)P\left(\mathcal{N}_i\right)\\ &\approx P\left(x_i^{(j)}|\mathcal{N}_i\right),
\end{split}
\end{align}
where the axion hypothesis $\mathcal{A}_i(g_\gamma)$ is here written explicitly as a family of hypotheses parametrized by $g_\gamma > 0$. The case of $g_\gamma=0$ is indistinguishable from the no-axion hypothesis $\mathcal{N}_i$. The approximation in Eq.\,\eqref{law tot prob} is justified so long as the total probability of there being an axion in any bin $i$, 
\begin{equation}
    P\left(\mathcal{A}_{i,\text{tot}}\right) = \int_0^\infty P\left(\mathcal{A}_i(g_\gamma)\right)dg_\gamma,
\end{equation}
is considerably smaller than any prior update that would be applied according to Eqs.\,\eqref{BPM update} and \eqref{BPM update tot}. 

To make a coarse estimate of $P\left(\mathcal{A}_{i,\text{tot}}\right)$, we can write down a Drake-like equation for axionic dark matter to which a typical haloscope searching for QCD axions such as HAYSTAC or ADMX might be sensitive: 
\begin{equation}\label{Drake}
    P\left(\mathcal{A}_{i,\text{tot}}\right) =  P(\text{PQ}) \times P(\text{DM}) \times P(m_a\approx h \nu_i /c^2),
\end{equation}
where $\text{PQ}$ denotes the event of the Peccei-Quinn hypothesis \cite{peccei1977cp, peccei1977constraints} being correct and the QCD axion field existing, $\text{DM}$ denotes the event of that axion field actually accounting for an appreciable fraction of the galaxy's dark matter \cite{preskill1983cosmology, abbott1983cosmological, dine1983not, turner1986cosmic}, $m_a\approx h \nu_i /c^2$ denotes the event that the axion is within roughly a linewidth of the bin of interest, $i$, and the probability of each event in parentheses is implicitly conditional upon all leftward events. In practice several other propositions must be true for an axion to be detectable. For instance, the axion must not be coincident with any other rf spikes and its lineshape should roughly match that used in the data processing \cite{brubaker2017first, du2018search}. 

It is beyond the scope of this article to attempt serious estimates of these probabilities, but for the sake of justifying the infinitesimal prior approximation, we only need to consider the sample optimistic set of probabilities shown in Table \ref{tab:Drake probs}. For logarithmically uniform distributed priors \cite{chaudhuri2018fundamental}
\begin{equation} \label{log priors}
    P_{a,i} \propto \frac{1}{\nu_i}
\end{equation}
in a $Q_a \sim 10^6$ axion between $m_\text{SN} \sim 1$ $\mu$eV and $m_\text{OC} \sim 1$ meV, the approximate mass range not disfavored by evidence from the SN1987a neutrino burst \cite{raffelt2008axions} or overclosure arguments \cite{preskill1983cosmology, abbott1983cosmological, dine1983not}, $P(m_a\approx h \nu_i /c^2) = \log(1+Q_a)/\log(m_\text{SN}/m_\text{OC}) = 1.4\times 10^{-7}$. If overclosure arguments do not apply \cite{graham2013new}, then a QCD axion could be far lighter than $m_\text{OC}$. Moreover, in practice, the prior probability for an axion to be in any given narrow coupling window must be less than $P(\mathcal{A}_{i,\text{tot}})$. Hence, the rough estimate of $P(\mathcal{A}_{i,\text{tot}})\sim 6\times 10^{-8}$ should be considered a generous upper bound on the prior probability as the term is used in this article. 

\begin{table}[]
    \centering
    \begin{tabular}{|l|l|} 
    \hline
    \textbf{event}& \textbf{optimistic probability} \\ \hline
    $\text{PQ}$ & $0.8$ \\ \hline
    $\text{DM}$ & $0.5$ \\ \hline
    $m_a\approx h \nu_i /c^2$ & $1.4\times10^{-7}$ \\ \hline
    $\mathcal{A}_{i,\text{tot}}$ & $6\times10^{-8}$ \\ \hline
    \end{tabular} 
    \caption{Sample set of probabilities used to estimate prior probability of an axion according to Eq.\,\eqref{Drake}. The probabilities $P(\text{PQ})$ and $P(\text{DM})$ are set to high (optimistic) values, such that raising them would not qualitatively change the conclusion that the priors are in the appropriate infinitesimal limit. The probability of the axion mass coinciding with a given bin assumes logarithmic priors \cite{chaudhuri2018fundamental}, and is itself optimistic in assuming that the axion exists within a $3$ mass decade window in which the haloscope operates. 
    } \label{tab:Drake probs}
\end{table}

In practice, one should not wait for a prior update of order $1/P(\mathcal{A}_{i,\text{tot}})$ to manually interrogate a bin;\footnote{Manual interrogation means performing tests that would discriminate between most spurious rf excesses and an axion: for example tuning the spike well off cavity resonance and/or ramping down the magnet and seeing if the signal persists.\label{FN: manual}} instead, an approximate prior estimate of the prevalence of spurious rf tones (Appendix \ref{app:non-Gauss}) should inform decision making. The prevalence of these rf spikes is \textit{a priori} unknown, but the fact that in practice \cite{du2018search, zhong2018results, brubaker2017haystac} the number of rescans required agrees with the predictions from Gaussian statistics (i.e. the measured distribution of excess powers looks Gaussian even in its right tail) indicates that only a few rf spikes large enough to push a bin above threshold are expected in a scan. For an initial scan with $N_I$ independent bins, therefore, a prior update of $U \sim N_I/N_\text{rf}$ should be sufficient cause to manually interrogate the bin. Here, $N_\text{rf}$ is the experimentalist's best guess for the expected number of real rf spikes in the scan window. Insofar as this guess is incorrect, the haloscope search will take longer than would be optimal. Rescanning and eliminating spurious rf tones ensures that the bias in the updates remains small, as well as conservative.\footnote{The conservative bias discussed in Appendix \ref{app:non-Gauss} results from interpreting the no axion hypothesis, $\mathcal{N}_i$ in Eq.\,\eqref{law tot prob}, to incorporate spurious rf tones. This makes $P\left(x_i^{(j)}|\mathcal{N}_i\right)$ larger in reality than as calculated from Eq.\,\eqref{pdf}. Using a slightly undersized denominator in Eq.\,\eqref{BPM update} results in a slight anti-exclusion bias.} 

The prior probability $P(\mathcal{A}_{i,\text{tot}})$ estimated in this appendix makes our approximation of infinitesimal priors valid for any single-bin updates $U\ll 1/P(\mathcal{A}_{i,\text{tot}})$. Since the prior update at which manual interrogation ought to begin, $N_I/N_\text{rf}$, is generally orders of magnitude less than  $1/P(\mathcal{A}_{i,\text{tot}})$, the approximation of infinitesimal priors will always be valid. 

\section{EFFECTS OF DATA PROCESSING ON INFORMATION CONTENT} \label{app:combine}
The main text of this article takes as its motivation the unacceptably long scan times that will likely be required to verify or falsify the axionic dark matter hypothesis over any meaningful fraction of the plausible parameter space. However, our framework optimizes only the statistical analysis of the already-processed data of a specific measurement, raising two complementary questions: is the measurement optimal, and is the data processing optimal?  The first question, of whether the measurement itself is optimal --- i.e. is the haloscope as typically operated the most efficient tool for evaluating the axionic dark matter hypothesis? --- is beyond the scope of this article, and we refer the reader to Ref.\,\cite{chaudhuri2018fundamental}. Subject to a haloscope or haloscope-like search platform, however, this appendix addresses the effect of the data processing --- the intermediate set of steps between measurement and statistical analysis --- on the information content of the haloscope data. We show that the current data processing protocol of the leading haloscope experiments \cite{du2018search, zhong2018results} are already highly optimized, with very little room for further improvements, though this has not historically been the case. 

We define the data processing as any numerical manipulation of a measured quantity from the moment of its digitization until it is put into a statistical analysis framework \cite{brubaker2017haystac, asztalos2001large}. The processing of real haloscope data involves many steps for dealing with practical nonidealities that are out of the scope of this appendix. We do not address issues such as the identification and removal of persistent but nonaxionic spectral features, or the digital high-pass filtering of the data. For a thorough treatment of the HAYSTAC experiment's processing protocol\footnote{The HAYSTAC processing protocol is in its essentials the same as that presently used by ADMX \cite{du2018search}.} including all these steps, see Ref.\,\cite{brubaker2017haystac}. This appendix evaluates only the mathematical operations as performed on an ideal dataset where the goal is to determine the presence or absence of an axion at each frequency. 

The data that may contain an axionic imprint arrives as a pair (one for each quadrature)\footnote{It is theoretically equally efficient to perform single-quadrature measurement \cite{malnou2019squeezed}, in which case there is only one quadrature time stream, but none of the conclusions of this appendix are altered.} of discretely sampled, identically distributed time-streams of measured voltages. Many such time-streams are obtained, and they are broken up and discretely Fourier transformed\footnote{While in principle it is possible to evaluate the axionic dark matter hypothesis in the time domain, we do not expect that this would provide any benefit, given the unitarity of the Fourier transform and the fact that the phase of the axion field at any moment is unknown.} into so-called subspectra, with frequency bin spacing $\Delta_b$ roughly two orders of magnitude narrower than the axion linewidth $\Delta_a$. The subspectra are acquired in batches of $N_b$ at each of $N_t$ distinct haloscope tuning steps. The Fourier frequencies in each subspectrum can be mapped faithfully back to the $N_f$ rf frequencies\footnote{The number of rf frequencies in earlier processing steps $N_f$ is generally larger than $N$, the number of frequencies in the grand spectrum, as a later data processing step will average sets of spectrally adjacent bins.} of the fluctuations within the haloscope cavity. Ultimately, each rf frequency is associated with a single bin in each of a large number of distinct subspectra. Therefore, the complex subspectrum voltages are denoted $v^\alpha_{Q, i j k}$, where $Q = X$ or $Y$ denotes the quadrature, $i=1,\dots,N_f$ indexes the rf frequency, $j =1,\dots,N_t$ the haloscope tuning step, and $k = 1,\dots,N_b$ the subspectrum at a given tuning step. 

At each of these tuning steps, sensitivity data are acquired as well. The sensitivity data, in a simplified picture, fully specify the expected thermal noise and possible axion signal power at each frequency bin $i$. At each tuning step and for each frequency, the sensitivity parameters $\eta^\gamma_{i j}$, defined as the ratio of signal power to noise power for a axion with $g_\gamma = 1$ that delivers $100\%$ of its power at the $i^\text{th}$ rf frequency, can then be calculated.

This appendix evaluates the consequences of each data processing step on the axion-pertinent (AP) information content of the data, where AP is understood to mean ``potentially bearing on the probability of the axion hypothesis being true." Using the terminology of Ref.\,\cite{brubaker2017haystac}, the subspectrum data are processed ultimately into grand spectrum data through the following steps:

\begin{enumerate}
\item \label{step: reim2pow} The real and imaginary parts of each subspectrum quadrature voltage are summed in quadrature so as to obtain the quadrature powers $x^\alpha_{Q, i j k}$ within each bin:
\begin{equation} \label{proc 1}
x^\alpha_{Q, i j k} = (\Im[v^\alpha_{Q, i j k}])^2 + (\Re[v^\alpha_{Q, i j k}])^2.
\end{equation}

\item \label{step: quads tot} The quadrature powers are summed to obtain the total power $x^\alpha_{i j k}$ in each subspectrum bin: 
\begin{equation} \label{proc 2}
x^\alpha_{i j k} = x^\alpha_{X, i j k} + x^\alpha_{Y, i j k}.
\end{equation}

\item \label{step: sub2raw} All subspectrum powers at each tuning step and frequency are summed to obtain the raw spectrum powers:
\begin{equation} \label{proc 3}
x^\beta_{i j} = \sum_{k = 1}^{N_b}x^\alpha_{i j k}.
\end{equation}

\item \label{step: raw2proc} The expected noise power, obtained from the application of a Savitzky-Golay (SG) filter \cite{savitzky1964smoothing, malagon2014search} at each frequency, is subtracted from each raw spectrum to obtain the processed spectra: \begin{equation} \label{proc 4}
x^\gamma_{i j} = x^\beta_{i j} - \braket{x^\beta_{i j}}.
\end{equation}
The processed spectra are power excesses: that is, departures of measured powers from the values expected absent an axion.

\item \label{step: proc2resc} The processed spectra are divided by the independently obtained sensitivities to obtain the rescaled spectra: 
\begin{equation} \label{proc 5}
x^\delta_{i j} = x^\gamma_{i j} / \eta^\gamma_{i j}.
\end{equation}
The rescaled spectra have the property that an axion with $g_\gamma = 1$ unrealistically depositing all of its power entirely within any one bin shifts the mean power excess for that bin up from 0 to 1. Less axion-sensitive bins are therefore drawn from distributions with larger variances $(\sigma^\delta_{ij})^2$. 

\item \label{step: resc2comb} In each rf frequency bin $i$, the combined spectrum power excess $x^\epsilon_{i}$ is taken as the ML-weighted sum of all rescaled spectra power excesses: 
\begin{equation} \label{proc 6}
\frac{x^\epsilon_{i}}{\left(\sigma_i^\epsilon\right)^2} = \sum_{j=1}^{N_t} \frac{x^\delta_{i j}} {\left(\sigma^\delta_{i j}\right)^2}.
\end{equation}
The variance of the $i^\textrm{th}$ combined spectrum bin is obtained from the variance of all contributing rescaled spectrum  bins as 
\begin{equation} \label{proc 6 var}
\left(\sigma_i^\epsilon\right)^2 = \left(\sum_{j=1}^{N_t} \frac{1}{\left(\sigma_{ij}^\delta\right)^2}\right)^{-1},
\end{equation}
where $(\sigma^\delta_{i j})^2 = \infty$ at tuning steps that do not contribute to a given frequency bin $i$.

\item \label{step: comb2reb} Sets of $n_c \ll \Delta_a/\Delta_b$ adjacent combined spectrum power excesses are rescaled, and the rebinned spectrum power excesses $x^\zeta_{i}$ are formed from their ML-weighted sums:
\begin{equation} \label{proc 7}
\frac{x^\zeta_{i}}{(\sigma^\zeta_{i})^2} = \frac{1}{n_c n_r}\sum_{l=1}^{n_c} \frac{x^\epsilon_{n_c (i-1)+l}}{\left(\sigma^\epsilon_{n_c (i-1)+l}\right)^2},
\end{equation}
where only every $n_c^\text{th}$ rebinned power excess is calculated and $n_r$ is the number of rebinned bins that contribute to each grand spectrum bin (Step \ref{step: reb2grand}). The rebinned variances are the analogously scaled, ML-estimated variances given by their combined spectrum counterparts:
\begin{equation} \label{proc 7 var}
(\sigma^\zeta_{i})^2 = \frac{1}{(n_c n_r)^2}\left(\sum_{l=1}^{n_c} \frac{1}{\left(\sigma^\epsilon_{n_c (i-1)+l}\right)^2}\right)^{-1},
\end{equation}

The rebinned bin frequencies are taken as the average of the contributing combined bin frequencies. For the rebinned spectrum and the subsequent grand spectrum, $i$ now indexes frequency in the more sparsely populated space. 

\item \label{step: reb2grand} The axion lineshape $f_a(\nu|\nu_a)$ given an axion with rest mass $\nu_a$ is calculated as the PDF of axion particle energies given their Maxwellian velocity distribution in the galactic rest frame, and our planet's velocity through that rest frame \cite{turner1990periodic}. This PDF is then discretized into a PMF $\bar{p}_a(\nu_j|\nu_{a,i})$ by integration over rebinned spectrum bins.\footnote{This conversion is made difficult by the fact that the axion frequency will in general not align perfectly with any bin frequency, an effect that matters less the smaller the rebinned bin spacing $n_c \Delta_b$ is. Reference \cite{brubaker2017haystac} accounts for this uncertainty.\label{FN: misalignment}} The grand spectrum is then constructed. The grand spectrum has the same frequency spacing as the rebinned spectrum, but each grand spectrum bin $i$ includes contributions from the $n_r = \Delta_a/n_c \Delta_b$ rebinned spectrum bins beginning at $i$. Sets of $2 n_r + 1$ grand spectrum bins are thus by construction correlated (Appendix \ref{app:corr}). The grand spectrum power excesses $x_i^\eta$ are constructed from the sets of $n_r$ adjacent rebinned power excesses as
\begin{equation} \label{proc 8}
\frac{x_i^\eta}{(\sigma^\eta_{i})^2} = \sum_{l=1}^{n_r} \frac{\bar{p}_a(\nu_{i + l - 1}|\nu_{a,i}) x^\zeta_{i + l - 1}}{\left(\sigma^\zeta_{i+l-1}\right)^2},
\end{equation}
where $\nu_{i+l-1}$ denote the rebinned frequencies. Equation \eqref{proc 8} provides the ML estimate of the power delivered by an axion to the $i^\text{th}$ grand spectrum bin, which has variance given by
\begin{equation} \label{proc 8 var}
    (\sigma^\eta_{i})^2 = \left(\sum_{l=1}^{n_r} \left[\frac{\bar{p}_a(\nu_{i + l - 1}|\nu_{a,i})}{\sigma^\zeta_{i+l-1}}\right]^2\right)^{-1}.
\end{equation}

\item \label{step: correct}
If the combined spectrum bins did not contain correlations from the SG filtering (Step \ref{step: raw2proc}), then the division by the independently obtained sensitivities (Step \ref{step: proc2resc}) would ensure that $x_i^\eta / \sigma^\eta_{i}$ is drawn from a standard normal distribution (no axion) or a normal distribution with unit mean and standard deviation (axion with $g_\gamma = 1$). The negative correlations imprinted by the SG filter (Appendix \ref{app:corr}), however, reduce the grand spectrum standard deviations $\sigma^\eta_{i}$ by a spectrally uniform factor $\xi$,\footnote{For phase 1 of HAYSTAC, $\xi = 0.93$ \cite{brubaker2017haystac}.} which can be extracted directly from the data and validated through simulation. 

Scaling the grand spectrum standard deviations down to 
\begin{equation} \label{proc 9 var}
    \tilde{\sigma}_i^\eta = \xi \sigma_i^\eta 
\end{equation}
makes the corrected grand spectrum excesses 
\begin{equation} \label{proc 9}
x_i^{(1)} = x_i^\eta / \tilde{\sigma}^\eta_{i}
\end{equation}
standard normal random variables absent an axion. The superscript $(1)$ denotes that Steps \ref{step: reim2pow}--\ref{step: correct} have applied for the initial scan, and must be repeated for rescans. Equation \eqref{proc 9} represents the grand spectrum excesses referred to in the main text. 

The SG filter has the additional effect of attenuating the visibility of a potentially present axion. The frequency-independent magnitude of this attenuation can be simulated as well, and is captured along with other, frequency-dependent effects, in Eq.\,\eqref{mu axion}'s sensitivity parameters $\eta_i^{(1)}$. 

\item \label{step: repeat} Steps \ref{step: reim2pow}--\ref{step: correct} are repeated, possibly multiple times, at frequencies where rescans are deemed appropriate.

\end{enumerate}

Information pertinent to the existence of the axion can only be generated during the measurement, and may be degraded by the subsequent data processing. Our aim is to determine where in Steps \ref{step: reim2pow}--\ref{step: repeat}, if at all, AP information could be lost. A guaranteed way to make use of all of the AP information content would therefore be to express the predictions of the axion and no-axion hypotheses, $\mathcal A_{Q,ijk}$ and $\mathcal N_{Q,ijk}$, respectively, in the space of the quadrature voltages $v^\alpha_{Q,ijk}$ and apply Bayes' theorem as 
\begin{equation} \label{voltage update}
u^{\alpha,\Re}_{Q,ijk} = \frac{P(\Re[v^\alpha_{Q,ijk}]|\mathcal A_{Q,ijk})}{P(\Re[v^\alpha_{Q,ijk}]|\mathcal N_{Q,ijk})},
\end{equation}
where the standard infinitesimal prior limit has been assumed and the same update applies for the imaginary components $\Im[v^\alpha_{Q,ijk}]$. While this approach is impractical, it nonetheless serves as an excellent baseline to compare the effects of the data processing protocol against. 

Step \ref{step: reim2pow} includes two operations: squaring the real and imaginary parts of each quadrature voltage, and then adding them. The real and imaginary parts of each quadrature voltage can be approximated as the same independent, identically distributed Gaussian random variables with mean zero \cite{clerk2010introduction, foster2018revealing}. Absent an axion, we label the variances of these random variables  $\sigma_{v,ij}^2$. An axion, assumed to be of a given coupling and at a given frequency, increases the variance to $\lambda_{ij} \sigma_{v,ij}^2,$ where $\lambda_{ij}$ is extremely close to but greater than unity. The prior update, Eq.\,\eqref{voltage update}, is the ratio of Gaussian PDFs given by Eq.\,\eqref{pdf}:
\begin{equation} \label{voltage update 2}
u^{\alpha,\Re}_{Q,ijk} = \frac{1}{\sqrt{\lambda_{ij}}}\exp\left[\frac{1-\lambda_{ij}^{-1}}{2\sigma_{v,ij}^2}\left(\Re[v^\alpha_{Q,ijk}]\right)^2\right].
\end{equation}
Since $\Re[v_{Q,ijk}]$ only occurs squared in this prior update (likewise $\Im[v_{Q,ijk}]$ in its identical equation), no AP information is lost by squaring the real and imaginary quadrature voltage components. Furthermore, since Bayesian updates are by their nature combined multiplicatively, the update accounting for both components of the quadrature voltage is
\begin{align}
\begin{split}\label{voltage update 3}
u^{\alpha,v}_{Q,ijk} = \frac{1}{\lambda_{ij}}&\exp\Bigg[\frac{1-\lambda_{ij}^{-1}}{2\sigma_{v,ij}^2}\\ &\left(\left(\Re[v^\alpha_{Q,ijk}]\right)^2+ \left(\Im[v^\alpha_{Q,ijk}]\right)^2\right)\Bigg].
\end{split}
\end{align}
Because the voltage quadrature components appear in the prior update only in their Eq.\,\eqref{proc 1} combination, Step \ref{step: reim2pow} preserves AP information content. 

The same argument applies to Steps \ref{step: quads tot} and \ref{step: sub2raw}. Multiplicative prior updates have their powers add in the exponent, hence Steps \ref{step: quads tot} and \ref{step: sub2raw} do not degrade AP information content. For Step \ref{step: quads tot}, this argument assumes that the axion and the thermal noise fluctuations distribute power evenly between the $X$ and $Y$ quadratures. For a two-quadrature measurement where the phase of the axion is unknown, that assumption is valid. 

Step \ref{step: raw2proc} is among the more vulnerable steps to meaningful information loss. If $\braket{x_{ij}^\beta}$ in Eq.\,\eqref{proc 4} is treated as a number, then Eq.\,\eqref{proc 4} amounts merely to the subtraction of a constant from the raw spectrum random variables $x_{ij}^\beta$, which will not degrade the information content. In practice, however, $\braket{x_{ij}^\beta}$ must itself be estimated, making it in effect a random variable. The estimation used by HAYSTAC and ADMX is performed via SG filters, which estimate $\braket{x_{ij}^\beta}$ at each frequency bin $i$ via the polynomial generalization of a moving average. If a wide spectral window for the generalized moving average is used, then $\braket{x_{ij}^\beta}$ is estimated with low variance, and acts like a constant. However, the SG filtering creates undesired correlations between bins up to two window-lengths apart, while also slightly attenuating an axion's visibility, an effect accounted for in Step \ref{step: correct}. The trade-off between the desired filtering effects and the undesired correlations and attenuation is beyond the scope of this article, but is discussed at length in Ref.\,\cite{brubaker2017haystac}. So far, no proof exists that SG is the optimal high-pass digital filter for the data, and we therefore identify Step \ref{step: raw2proc} as potentially admitting of meaningful optimization. 

Calculating the processed spectrum sensitivity parameters $\eta^\gamma_{ij}$ of Step \ref{step: proc2resc} relies on independent measurements of the thermal and added noise, magnetic field, and cavity quality factor and mode-structure properties of the haloscope, as well as calculations of the power that a dark matter axion would deliver. Subject to the accuracy of these measurements and calculations, Step \ref{step: proc2resc} simply divides the processed spectra power excess random variables $x_{ij}^\gamma$ by constants, preserving information content. 

The ML-weighting of Step \ref{step: resc2comb} is the information-preserving generalization of the straightforward addition of Steps \ref{step: reim2pow}--\ref{step: sub2raw}, for the case where the random variables being added have different variances. By the central limit theorem, the rescaled power excesses $x_{ij}^\delta$ are Gaussian-distributed with known variances $\sigma_{ij}^\delta$ well approximated as independent of the presence of any axion which delivers power far less than vacuum. The pertinent effect of an axion is to shift the mean of the rescaled power excess from $0$ to $\mu^{\delta}$, which has no bin- or tuning step-dependence by construction. The appropriate prior update delivered by each rescaled power excess $x_{ij}^\delta$ is thus
\begin{equation}\label{rescaled update}
u^\delta_{ij} = \exp\left[\frac{-(\mu^\delta)^2/2 + \mu^\delta x_{ij}^\delta}{(\sigma_{ij}^\delta)^2} \right]. 
\end{equation}
When all $N_t$ tuning steps' updates are multiplied, the result is
\begin{align}
\begin{split}\label{rescaled update product}
\prod_{j=1}^{N_t} u^\delta_{ij} &= \exp\left[\sum_{j=1}^{N_t} \left(\frac{-(\mu^\delta)^2/2 + \mu^\delta x_{ij}^\delta}{(\sigma_{ij}^\delta)^2} \right)\right]\\
&= \exp\left[\frac{-(\mu^\delta)^2/2 + \mu^\delta x_{i}^\epsilon}{(\sigma_{i}^\epsilon)^2} \right].
\end{split}
\end{align}
Because the updates from all bin-$i$ rescaled spectra power excesses $x_{ij}^\delta$ and variances $(\sigma_{ij}^\delta)^2$ can thus be obtained using the single bin-$i$ combined spectrum power excess $x_i^\epsilon$ and variance $(\sigma_i^\epsilon)^2$, given by Eqs.\,\eqref{proc 6} and \eqref{proc 6 var}, respectively, the ML estimation of Step \ref{step: resc2comb} preserves AP information content. 

The same argument about ML estimation guarantees that Steps \ref{step: comb2reb} and \ref{step: reb2grand} preserve AP information as well, in the limit where the axion lineshape is truly approximately constant over spectral scales of $n_c \Delta_b$. Step \ref{step: correct} is in effect properly accounting for the information loss suffered during Step \ref{step: raw2proc}, but is itself simply multiplication by a scalar, and causes no additional information loss. 

In summary, the use of ML estimation, not consistently applied in previous haloscope analyses \cite{asztalos2001large}, makes the grand spectrum excesses $x_i^{(j)}$ together with the sensitivity parameters $\eta_i^{(j)}$ nearly sufficient statistics for evaluating the cold dark matter axion hypothesis. We identify only two areas where the sensitivity can be improved. First, it is possible that a high-pass filter with lower stop-band attenuation that maintains acceptable pass-band performance would improve the processing sensitivity. Second, the rationale behind the rebinning (Step \ref{step: comb2reb}) --- that it will reduce the expected number of axion candidates exceeding threshold, mitigating the look-elsewhere effect --- is unnecessary in the Bayesian power measured framework. By not rebinning (setting $n_c \rightarrow n_c^\prime = 1$ and $n_r \rightarrow n_r^\prime = n_c n_r$), the curvature of the expected axion lineshape over $n_c \Delta_b$-sized spectral scales is optimally accounted for. The expected $n_c$-fold growth of such candidates introduced by not rebinning is precisely negated in the aggregate prior update, Eq.\,\eqref{agg update}, by the $n_c$-fold growth in the number $N$ of grand spectrum bins.\footnote{Future analyses may on the same principle set grand spectrum bin spacings closer than those of the combined spectrum. Doing so eliminates the need to account for misalignment (Footnote \ref{FN: misalignment}) of the axion rest mass with grand spectrum bin frequency.}

\section{ADDITIONAL FEATURES OF THE HAYSTAC PHASE 1 DATASET} \label{app:features}
In addition to the LU candidate at $\nu_{LU}^\star$ discussed in Sec.\,\ref{sec:HAYSTAC}, the dashed, blue curve of Fig.\,\ref{fig:HAYSTAC data}b also neglects two other features of the HAYSTAC phase 1 dataset that appear in Fig.\,\ref{fig:HAYSTAC data}a. First, the BPM reanalysis includes data from rescans over the upper 100 MHz of the scan window that were performed during phase 1, run 1 of HAYSTAC \cite{brubaker2017first}. These rescans were discarded, and new ones performed over some of the same frequencies, because an error in estimating the expected axion lineshape lowered their sensitivities below the levels required by the predetermined confidence levels and initial scan sensitivities \cite{brubaker2017haystac}. The extra rescan data itself is valid, however, and is thus included in Fig.\,\ref{fig:HAYSTAC data}a and b, showcasing the flexibility of BPM to integrate all available information. We do not include this extra data in the comparison (dashed blue) curve, as the existence of extraneous rescan data is not ordinary in haloscope analyses.

Secondly, a $3.94\sigma$ grand spectrum excess at grand spectrum bin 550 ($\nu_{550} = 5.59886155$) GHz was recorded in the initial scan. This excess occurs $0.3\%$ of the way into the scan window, where the number of contributing raw spectra (Appendix \ref{app:combine}) was 14, considerably less than the typical 40. The lower number of spectra implies at once an increased susceptibility to systematic error and a reduced axion sensitivity $\eta_{550}^{(1)} = 0.23$. The reduced sensitivity implies that an axion here has its prior update, Eq.\,\eqref{BPM update}, maximized at $g_\gamma \approx 4.1$. This excess ought to have been rescanned during phase 1 of HAYSTAC, but was not. As such, we include its positive prior update in the BPM reanalysis. Its presence is indicated by the dark column at the left of Fig.\,\ref{fig:HAYSTAC data}a, and is solely responsible for the failure of the aggregate prior update (solid blue line, Fig.\,\ref{fig:HAYSTAC data}) to vanish at high couplings $\gsim 3.2 g_\gamma^\text{KSVZ}$. Had the required rescans been performed, a source of excess would likely have been discovered or rejected in short order, as in the case of the LU candidate at $\nu_{LU}^\star$. As such, we remove it from the comparison (dashed blue) aggregate update curve.

\vspace{0.1in}

%


\end{document}